\documentclass[longauth]{aa}  
\usepackage{graphicx}
\usepackage{txfonts}
\usepackage[normalem]{ulem}

\usepackage{amsmath}

\usepackage{tabularx}
\usepackage{scrextend}
\usepackage{lscape} 
\usepackage{bigstrut}

\usepackage{textcomp} 

\usepackage{lineno} 

\usepackage{soul}
\usepackage{color}
\newcommand{\mr}[1]{{#1}}

\newcommand{\pkg}[1]{\textsc{\texttt{#1}}}

\newcommand{\sep}{\pkg{sep}}
\newcommand{\pysedm}{\pkg{pysedm}}

\newcommand{\pipeinput}[1]{\texttt{#1}}

\usepackage{natbib,twoopt}
\usepackage[breaklinks=true, draft]{hyperref} 
\bibpunct{(}{)}{;}{a}{}{,}             
\makeatletter
  \newcommandtwoopt{\citeads}[3][][]{\href{http://adsabs.harvard.edu/abs/#3}%
    {\def\hyper@linkstart##1##2{}%
     \let\hyper@linkend\@empty\citealp[#1][#2]{#3}}}
  \newcommandtwoopt{\citepads}[3][][]{\href{http://adsabs.harvard.edu/abs/#3}%
    {\def\hyper@linkstart##1##2{}%
     \let\hyper@linkend\@empty\citep[#1][#2]{#3}}}
  \newcommandtwoopt{\citetads}[3][][]{\href{http://adsabs.harvard.edu/abs/#3}%
    {\def\hyper@linkstart##1##2{}%
     \let\hyper@linkend\@empty\citet[#1][#2]{#3}}}
  \newcommandtwoopt{\citeyearads}[3][][]%
    {\href{http://adsabs.harvard.edu/abs/#3}
    {\def\hyper@linkstart##1##2{}%
     \let\hyper@linkend\@empty\citeyear[#1][#2]{#3}}}
\makeatother

\begin{document}

\title{A Fully Automated Integral Field Spectrograph Pipeline for the
  SEDMachine: pysedm}

\titlerunning{pysedm}
\authorrunning{M.~Rigault, J.~D.~Neill et al.}

\author{M.~Rigault \inst{\ref{lpc}},  J.~D.~Neill
  \inst{\ref{caltech}}, N.~Blagorodnova \inst{\ref{Radboud}},
  A.~Dugas \inst{\ref{caltech}}, M.~Feeney \inst{\ref{caltech}},  R.~Walters
    \inst{\ref{caltech}, \ref{caltechobs}}, Y.~Copin\inst{\ref{ipnl}},
    V.~Brinnel\inst{\ref{humboldt}}, C.~Fremling\inst{\ref{caltech}},
    J.~Nordin\inst{\ref{humboldt}}, J.~Sollerman \inst{\ref{okc}}
}


\institute{
    Université Clermont Auvergne, CNRS/IN2P3, Laboratoire de Physique
    de Clermont, F-63000 Clermont-Ferrand, France. \label{lpc}
\and
   Division of Physics, Mathematics, and Astronomy, California
   Institute of Technology, Pasadena, CA 91125, USA. \label{caltech}
\and
  Department of Astrophysics/IMAPP, Radboud University, Nijmegen, The
  Netherlands. \label{Radboud}
\and 
  Caltech Optical Observatories, California Institute of Technology,
  Pasadena, CA 91125, USA. \label{caltechobs}
\and
  Université de Lyon, F-69622, Lyon, France; Université de Lyon
  1, Villeurbanne; CNRS/IN2P3, Institut de Physique Nucléaire de
  Lyon. \label{ipnl}
\and
  Institute of Physics, Humboldt-Universität zu Berlin, Newtonstr. 15,
  124 89 Berlin, Germany. \label{humboldt}
\and
  The Oskar Klein Centre \& Department of Astronomy, Stockholm 
  University, AlbaNova, SE-106 91 Stockholm, Sweden. \label{okc}
}

\date{}

\abstract{

Current time domain facilities are discovering hundreds of new
galactic and extra-galactic transients every week. Classifying the
ever-increasing number of transients is challenging, yet crucial to
further our understanding of their nature, discover new classes, 
or ensuring sample purity, for instance, for Supernova Ia cosmology.
The Zwicky Transient Facility is one example of such a survey. In
addition, it has a dedicated very-low resolution spectrograph, the
SEDMachine, operating on the Palomar 60-inch telescope. This
spectrograph's primary aim is object classification. In practice most, if not 
all, transients of interest brighter than $\sim$19~mag are typed.
This corresponds to approximately 10 to 15 targets a night. 
In this paper, we present a fully automated pipeline for the
SEDMachine. This pipeline has been designed to be fast, robust, stable and
extremely flexible. \pysedm{} enables the fully automated spectral
extraction of a targeted point source object in less than 5 minutes after
the end of the exposure. The spectral color calibration is accurate at the
few percent level. In the 19 weeks since \pysedm{} entered
production in early August of 2018, we have classified, \mr{among other
objects}, about 400
Type Ia supernovae and 140 Type II supernovae. We conclude that low resolution,
fully automated spectrographs such as the ``SEDMachine with pysedm''
installed on 2-m class telescopes within the southern hemisphere could 
allow us to automatically and simultaneously type and obtain a redshift
for most (if not all) \mr{bright} transients detected by LSST within z<0.2,
\mr{notably potentially all Type Ia Supernovae. In comparison to the
  current SEDM design, this would require higher spectral
  resolution (R$\gtrsim$1000) and slightly improved throughput}. 
With this perspective in mind, pysedm has been designed to easily be
adaptable to any IFU-like spectrograph (see
https://github.com/MickaelRigault/pysedm).  
}


\maketitle

\section{Introduction}
\label{sec:introduction}

Time domain astronomy, the study of transients, variables and/or
moving objects, is one of the frontier fields of this decade. 
Modern surveys are now able to scan the entire visible sky with a daily
or near-daily cadence. The Zwicky Transient Facility
\citep[ZTF][]{ztf_paper, ztf_paper2}, with its \mr{47} deg$^{2}$ field of view and rapid camera, is one
example of such a survey. Other examples include the Catalina Real-Time Transient Survey
\citep{catalina_paper},
PanSTARRS-1 \citep{panstarrs_paper}, ASAS-SN \citep{assassin_paper} and
ATLAS \citep{atlas_paper}. While scanning the sky every night with a
typical 
$5\sigma$ magnitude limit of 20.5, ZTF detects of order 
$10^5$ variations in the sky between a new observation
and a past reference frame, generating an alert for each. While most of
these alerts are glitches or
known variable sources, about $\mathcal{O}(10^2)$ are transients of interest, such
as supernovae or any kind of galactic or extra-galactic
explosion. About 10\% of them are new transients that have just
exploded, or just become bright enough to be detected. Soon the Large
Synoptic Survey Telescope \citep[LSST,][]{lsst_paper}, with a
magnitude limit of 27.5 in r-band, will
detect 10 times more transients than ZTF, resulting in hundreds of new
extra-galactic events every night. 

In that context, classifying the ever-increasing number of new
supernovae (or equivalent) is a technical challenge. Yet a good
classification is crucial to further our understanding of the nature of
these transients, \mr{to investigate} their rates and to detect new
classes of objects.  For cosmology,
an accurate classification is required to ensure that the Hubble diagram is not
contaminated by non-Type Ia
supernovae. A major effort is underway to develop photometric
typing \citep[e.g.][and references therein]{sako_2008,ishida_2013,
  lochner_2016, moller_2016}. However, these methods have not 
yet proven to be sufficient, notably in cosmology, where non-Ia
contamination could significantly bias derivations of the cosmological
parameters \citep[e.g.][]{jones_2017}. In addition, without a spectral 
identification, new object families could be left undiscovered. 

In \cite{sedm_paper} we first presented the SEDMachine (SEDM), a
very low-resolution (R$\sim$100) spectrograph designed and built to type
transients discovered by ZTF. Since the
beginning of this survey in early June 2018, ZTF has discovered and
classified over 1000 new extra-galactic transients, most of them
Type Ia supernovae with $z<0.15$ \cite{feindt_2019}. The SEDM
demonstrates that typing a large fraction (if not all brighter than a 
certain magnitude) of detected
nearby transients is doable with dedicated facilities, \mr{typically
$\lesssim 20\,\mathrm{mag}$ if} installed on 1-2m class telescopes. 
For such a facility to be efficient, however, two key elements
are necessary: full automation and high reliability. While
currently the selection of transients sent to the SEDM is only partially
automated, since early August 2018 the new SEDM pipeline, \pysedm{},
has been in production and has extracted all submitted transient
spectra automatically.  

In this paper we present \pysedm{}. 
This software is meant to be a generic pipeline that could easily be
adapted to any further SEDM-like instrument. The code is public and
written in Python\footnote{https://github.com/MickaelRigault/pysedm}.  
It has not been designed to optimize the spectrophotometry, but rather
to be fast, robust, fully automated and easily customizable. 

\begin{figure*}
  \centering
  \includegraphics[width=\linewidth]{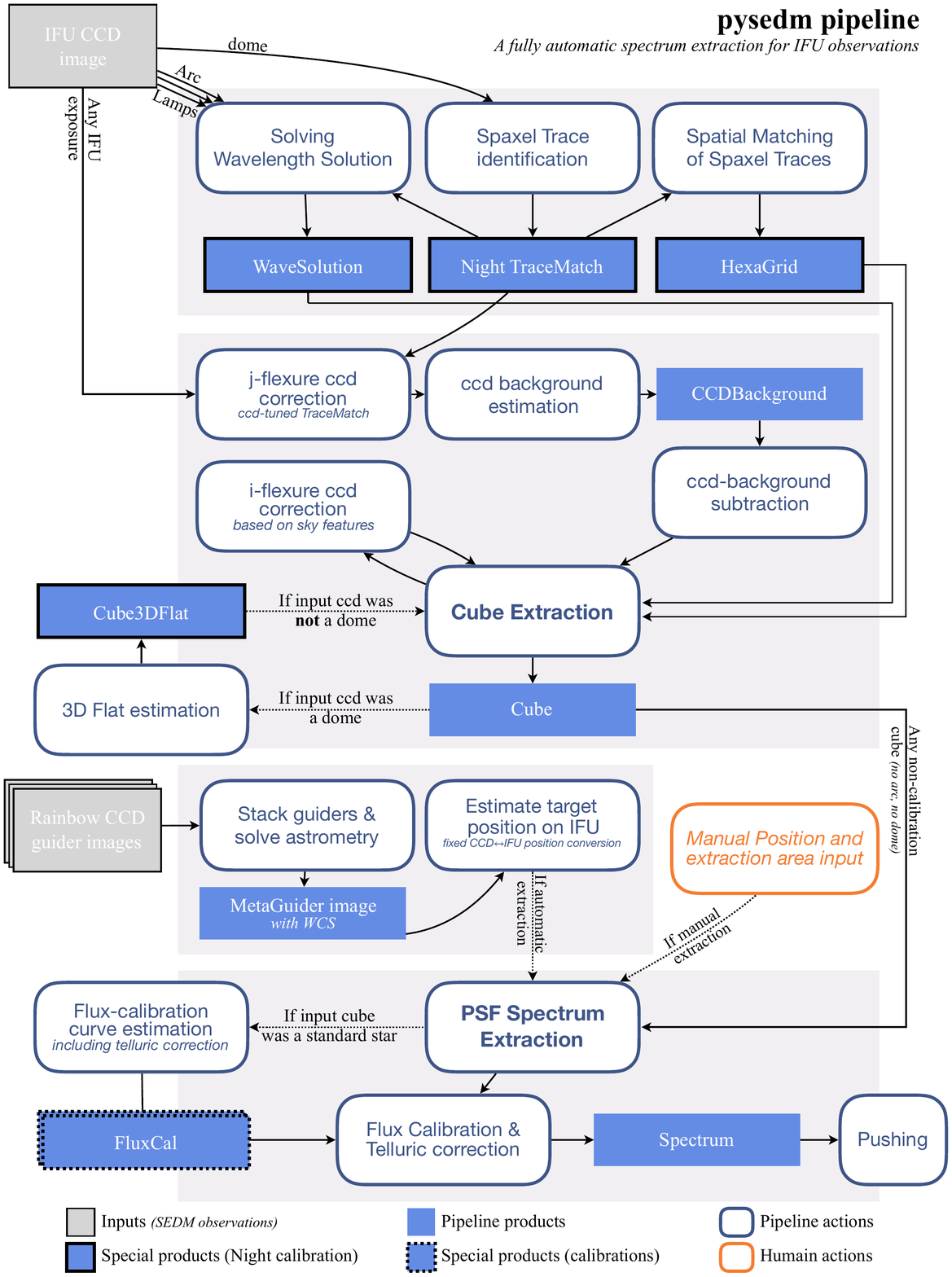}
  \caption{\pysedm\ pipeline}
  \label{fig:pysedmpipeline}
\end{figure*}

The pipeline is illustrated in
Fig.~\ref{fig:pysedmpipeline}. Section~\ref{sec:cube_extraction} we
explain the cube extraction from the raw  
observations of the nightly calibration solutions including the trace
identification (section~\ref{sec:nightcalibration_tracematch}), and
the wavelength solution (section~\ref{sec:nightcalibration_wavesol}).
Then, we present in section~\ref{sec:extract_star} the spectral
extraction from the 3D cubes.
An overall pipeline performance and some results are presented in
section~\ref{sec:results}. We conclude in section \ref{sec:conclusion}.

\section{Extracting a 3D Cube}
\label{sec:cube_extraction}

In integral field spectroscopy, extracting a 3D-cube from a 2D-image
requires the identification of three elements: (1) where spaxel light is
dispersed onto the CCD --~forming a ``trace''~--, (2) the correspondance
between a trace's location on the CCD and the position of the spaxel in the
focal plane and (3) the spectral dispersion mapping that converts each
CCD-pixel into actual wavelengths.

In the implementation of \pysedm{} for the SEDM, these elements are derived
once for the entire night based on dome and arc-lamp observations taken
during late afternoon. During nighttime observations, instrumental flexure
is solved for using night sky spaxels, and each image is aligned to the
nightly calibration solution.  This approach enables us to maximize the
time dedicated to science exposures, as no further calibration observations
are required. In addition, since the calibration steps are made during
daytime, the \pysedm\ pipeline can extract an automatic target spectrum
within 5 min after the end of the exposure.

In this section we detail the cube-extraction mechanism. We start by
presenting the night calibration solutions  in
section~\ref{sec:nightcalibration}, then in section~\ref{sec:ccd_to_cube}
we explain how to use them to extract the 3D cubes.

\subsection{Night Calibration}
\label{sec:nightcalibration}

Here we present the details of the algorithmic solutions developed in
\pysedm\ to build the calibrations representing the three aforementioned
elements: ``Trace Matching'', the ``Spatial Identification'',  and the
``Wavelength Solution''. An illustration of a raw SEDM CCD image is shown in
fig.~\ref{fig:tracematch}; here a dome flat. We see that spaxel traces are dispersed 
nearly horizontally, with bluer wavelengths toward the right of the traces.

\subsubsection{Trace Matching}
\label{sec:nightcalibration_tracematch}
\begin{itemize}
\item Input: \pipeinput{dome.fits} ;
\item Output: \pipeinput{TraceMatch.fits}
\end{itemize}

\begin{figure}
  \centering
  \includegraphics[width=\linewidth]{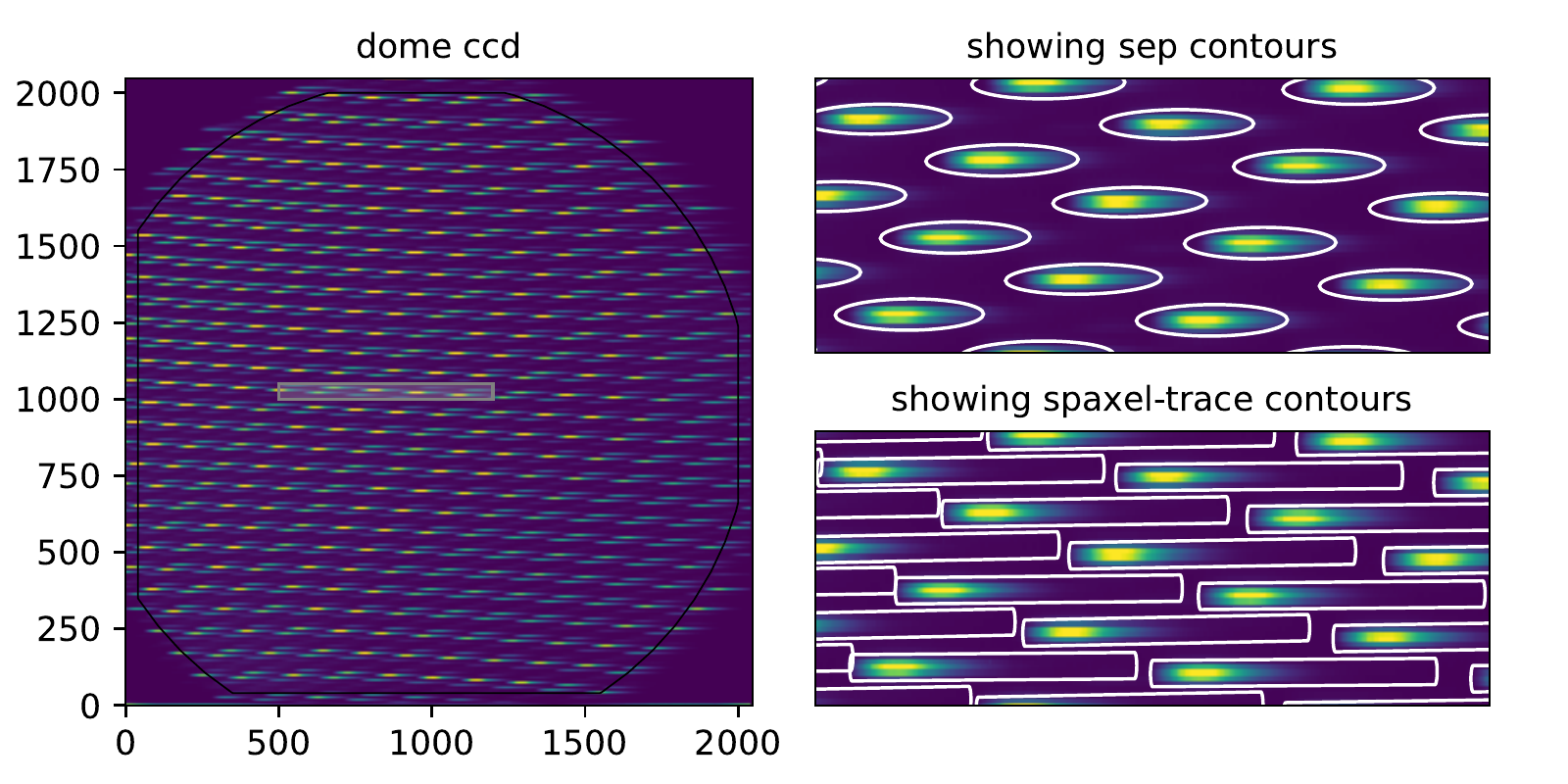}
  \caption{Extracting the trace contours from a SEDM dome
    exposure. \textit{Left} CCD image of
    a dome exposure. The grey rectangle shows the zoom area displayed
    on the right. \textit{Upper right}: \sep{} ellipses overplotted on
  the CCD dome image. \textit{Lower right}: Spaxel-trace contours overplotted on
  the CCD dome image. For the SEDM, spaxel-trace contours are defined as tilted
  rectangles. The rotation angle of the centroid position are derived
  from the \sep{} ellipses.}
  \label{fig:tracematch}
\end{figure}

To identify each individual spaxel trace, we run
the ``extract'' method of \sep\footnote{http://github.com/kbarbary/sep ;
  version used: 1.0.1} 
\citep{barbary_2016} --~the Python
implementation of \pkg{Sextractor} \citep{bertin_1996}~-- on a dome
CCD-exposure.
The \sep\ function defines an ellipse contour on dispersed dome light
for each spaxel; see Fig.~\ref{fig:tracematch}.
In the \pysedm\ implementation for the SEDM, spaxel traces are
defined as the rectangles formed by taking a 2.4 pixels region above and
below the spaxel ellipse centroid --~so 5.8 pixels in total~-- and
starting 70 x-pixels to the left of the ellipse centroid and ending
230 x-pixels to the right ; \mr{the dome light times the instrument
  response been mostly red}. 

The 2.4 pixel width has been selected to encapsulate 95.5\% ($\pm 2\sigma$)
of the spaxel light given the SEDM cross dispersion. As shown in
Fig.~\ref{fig:tracematch}, the compactness of spaxel traces in our 
instrument does not allow for significantly larger width.

The trace extent of 70 pixels to the left and 230 pixels to the right has
been chosen to be the smallest length that always encapsulates wavelengths
from 3500 to 9500 $\AA$, after wavelength solution conversion (see
section~\ref{sec:nightcalibration_wavesol}).

Once a spaxel trace is defined, we can create its corresponding 2D CCD
weight-mask defined such that each pixel corresponds to the fraction of its
area that lies within the trace contours; 0 for outside, 1 for fully inside, and intermediate values for pixels on an edge. 
This fraction is measured using
\pkg{shapely}\footnote{https://pypi.org/project/Shapely ; version
  used: 1.6.3}.

Given this 2D mask and because the spectral dispersion of the SEDM is
almost horizontal, one can extract a spaxel spectrum in counts per pixel
(named ``pixel-spectrum'') by summing the product ``CCD-image $\times$
spaxel weight-mask'' along the vertical-axis (``j-axis'').  For
convenience, when extracting a pixel-spectrum we count the pixels from
right to left such that increasing pixels correspond to increasing
wavelengths.

\subsubsection{Wavelength Solution}
\label{sec:nightcalibration_wavesol}

\begin{itemize}
\item Input:
  \begin{itemize}
  \item[$\bullet$] \pipeinput{Hg.fits}, \pipeinput{Xe.fits}, \pipeinput{Cd.fits}
  \item[$\bullet$] \pipeinput{TraceMatch.fits} (see Section~\ref{sec:nightcalibration_tracematch}) ;
  \end{itemize}
\item Output: \pipeinput{WaveSolution.fits}.
\end{itemize}

\begin{figure}
  \centering
  \includegraphics[width=\linewidth]{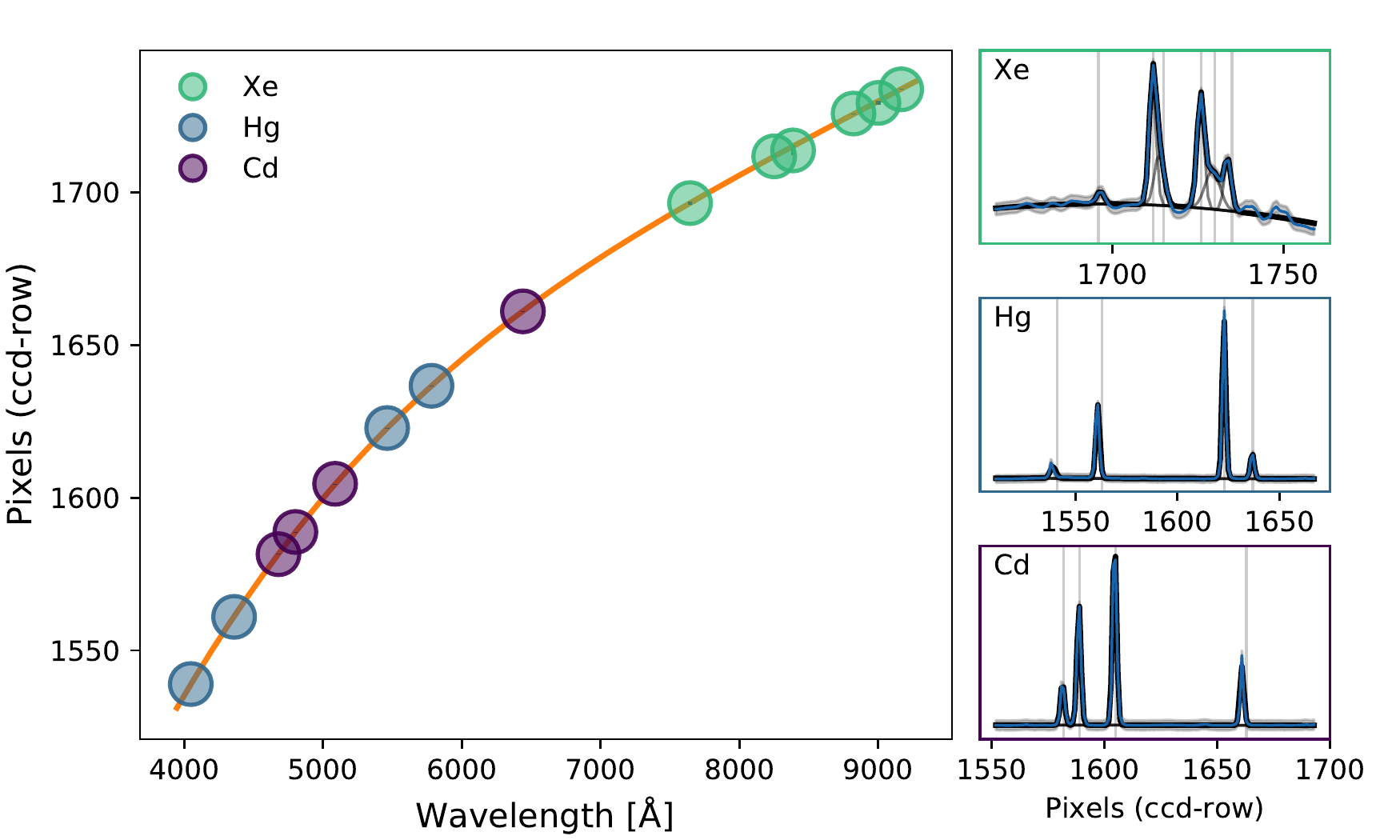}
  \caption{Wavelength solution fit for spaxel \#501.
    \textit{Left}: Fitted centroids of Hg, Cd and Xe (see legend) arc lamp emission
    lines (in pixel-row from the right) as a function of their
    expected wavelengths. The orange line is the best
    fitted 5$^\mathrm{th}$ order polynomial enabling the conversion of wavelength
    to pixel and conversely.
  \textit{Right}: Blue curves represent the arc lamp spectra (in flux per
  pixel-row from the right). The black lines are the best fitted
  ``Gaussians + third-order polynomial-continuum''. Vertical
  thin grey lines are the initial guess positions. See details in
  section~\ref{sec:nightcalibration_wavesol}.}
  \label{fig:wavesolution}
\end{figure}

A wavelength solution is derived for each spaxel independently based
on arc-lamp observations. We use three lamps for the SEDM
wavelength calibration (wavelength in vacuum); see fig.~\ref{fig:wavesolution}:
 \begin{itemize}
    \item[Hg:] A mercury lamp (Hg) that has four strong emission lines
      at 4047.7, 4359.6, 5462.3 and  5781.7 $\AA$ (a blend of 5771.2
      and 5792.3). 
\item[Cd:] A cadmium lamp (Cd) that has four strong emission lines
      at 4679.3, 4801.3, 5087.2 and 6440.2 $\AA$.
\item[Xe:] A xenon lamp (Xe) that has six emission lines
      at 7644.1, 8250.1 (a blend of 8233.90 and 8282.39),
      8386.2 (a blend of 8349.1 and 8411.0), 8821.8, 9001.3 (a blend of
      8954.7 and 9047.9) and 9165.1 $\AA$. The bluer Xe emission line
      is faint, but we decided to use it nonetheless to minimize the gap
      around 7500~$\AA$; see fig.~\ref{fig:wavesolution}.
\end{itemize}

For a given spaxel, we extract the corresponding pixel-spectrum (see
section~\ref{sec:nightcalibration_tracematch}) for each of the
three arc lamps and independently fit them with a combination of
Gaussian lines plus a third-order polynomial continuum.. The
Xe fit uses six Gaussians while the Cd and Hg fits only have four.
In a second iteration, we jointly fit the bijective relation between the 14
centroids (in pixels) as a function of their expected wavelengths with
a fifth-order polynomial.
The best fitted polynomial is recorded and defines the wavelength
solution for the given spaxels. An illustration of the procedure is
shown in fig.~\ref{fig:wavesolution}.

\begin{figure}
  \centering
  \includegraphics[width=0.7\linewidth]{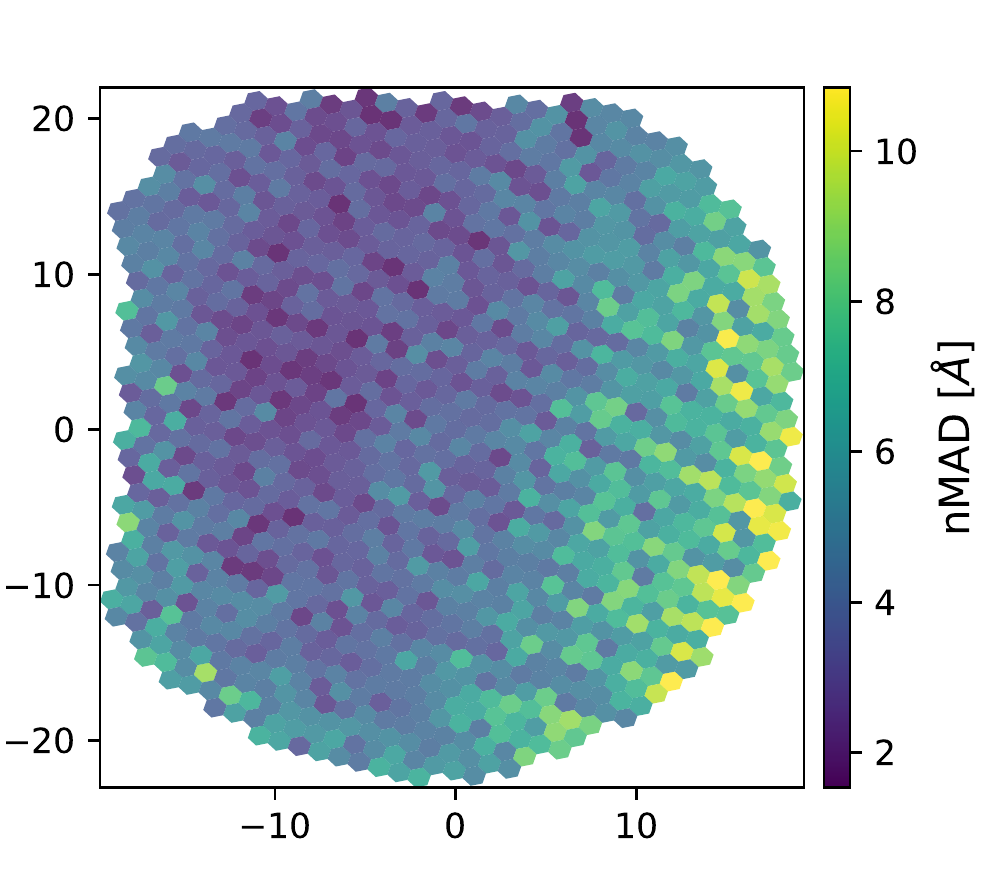}
  \caption{Illustration of the precision of the wavesolution
    procedure. The color of each spaxel corresponds to the normalized
    median average distance (nMAD) of measured arc lines around the
    wavelength solution. As discussed in the text, the meaningful
    information is the relative precision between spaxels: our
    wavelength solution is better toward the center of the IFU, where
    signal to noise is higher and science targets are designed
    to be located.}
  \label{fig:wavesolutionmap}
\end{figure}

By construction, each spaxel wavelength solution is independent. As a
validation plot for this calibration step, we record the normalized median
average distance (nMAD) around the wavelength solution for each spaxel; a
typical example is shown in Fig.~\ref{fig:wavesolutionmap}.  This scatter plot
contains both Gaussian line fit noise and potential inaccuracy in the
``true wavelength''  definition for the  arc lamps, especially for the
blended lines.  Hence, while the absolute nMAD value might not be so
relevant, the relative precision is meaningful and we see in
fig.~\ref{fig:wavesolutionmap} that the wavelength solution is more precise
in the center of the micro-lens array (MLA) than on its right edge (see details on spatial
solution in section~\ref{sec:nightcalibration_hexagrid}). If a wavelength
solution fails for a given spaxel, its nMAD would be much larger than a few
tens of \"angstroms. While this so far has not happened as long as the pipeline
has been in production, such a figure would enable us to identify any potential
problems with the wavelength solution calibration step. \mr{A test of
the wavelength calibration precision is presented in
Section~\ref{sec:test_wavesolution}.}  

\subsubsection{Spatial Identification}
\label{sec:nightcalibration_hexagrid}
\begin{itemize}
\item Input: \pipeinput{TraceMatch.fits} ;
\item Output: \pipeinput{HexaGrid.fits}.
\end{itemize}

The SEDM has a hexagonal MLA located in the focal plane
of the Palomar 60" (P60) telescope. The SEDM therefore creates a fully-filled hexagonal
$x,y,\lambda$ cube. The task of the spatial identification step is to
recover this hexagonal structure given the spaxel dispersion trace location
on the CCD. This is done in eight steps.

First, we use the spaxel-trace centroid to define spaxel reference
positions on the CCD. We use the trace-id from
\pipeinput{TraceMatch.fits} to identify the centroids.

Second, we integrate the centroids into a k-d tree and associate each
spaxel with its six nearest neighbors. Given the hexagonal structure of the
MLA, these are the six contiguous lenses.

Third, we start the hexagonal reconstruction around a ``reference spaxel'',
set at the beginning to the most central centroid on the CCD. We then pick
the nearest of the six neighbors to form a ``spaxel-pair''.  By
construction, this pair only has two shared neighbors, except if this pair
is located at the edge of the MLA in which case it only has one.

Fourth, we select the closest of the two shared neighbors of the spaxel-pair
to form a triangle. This first such triangle defines the \{q, r\}
coordinates of the hexagonal structure.

Fifth, we form a new ``spaxel-pair'' by pairing the reference spaxel with the
third spaxel of the triangle. By construction, this pair only has at
maximum one unknown neighbor, since we already know they have the
``second-spaxel'' in common. If it exists, i.e., if the pair is not at the
edge, we record the \{q,r\} coordinates of this fourth centroid. At this
stage, the reference spaxel and the third spaxel connect with three of
their neighbors, while the fourth and the second connect to only two.

Sixth, we repeat the fifth step updating the pairing with the reference
spaxel to the latest associated spaxel. After three iterations we can
connect all the six reference spaxel neighbors, and each of these are
already connected to three of their six neighbors.

Seventh, we repeat the fifth and sixth steps by randomly selecting one of
the six neighbors of the reference spaxel to become the new reference. We
continue until all spaxel centroids have been assigned.

\begin{figure*}
  \centering
  \includegraphics[width=0.7\linewidth]{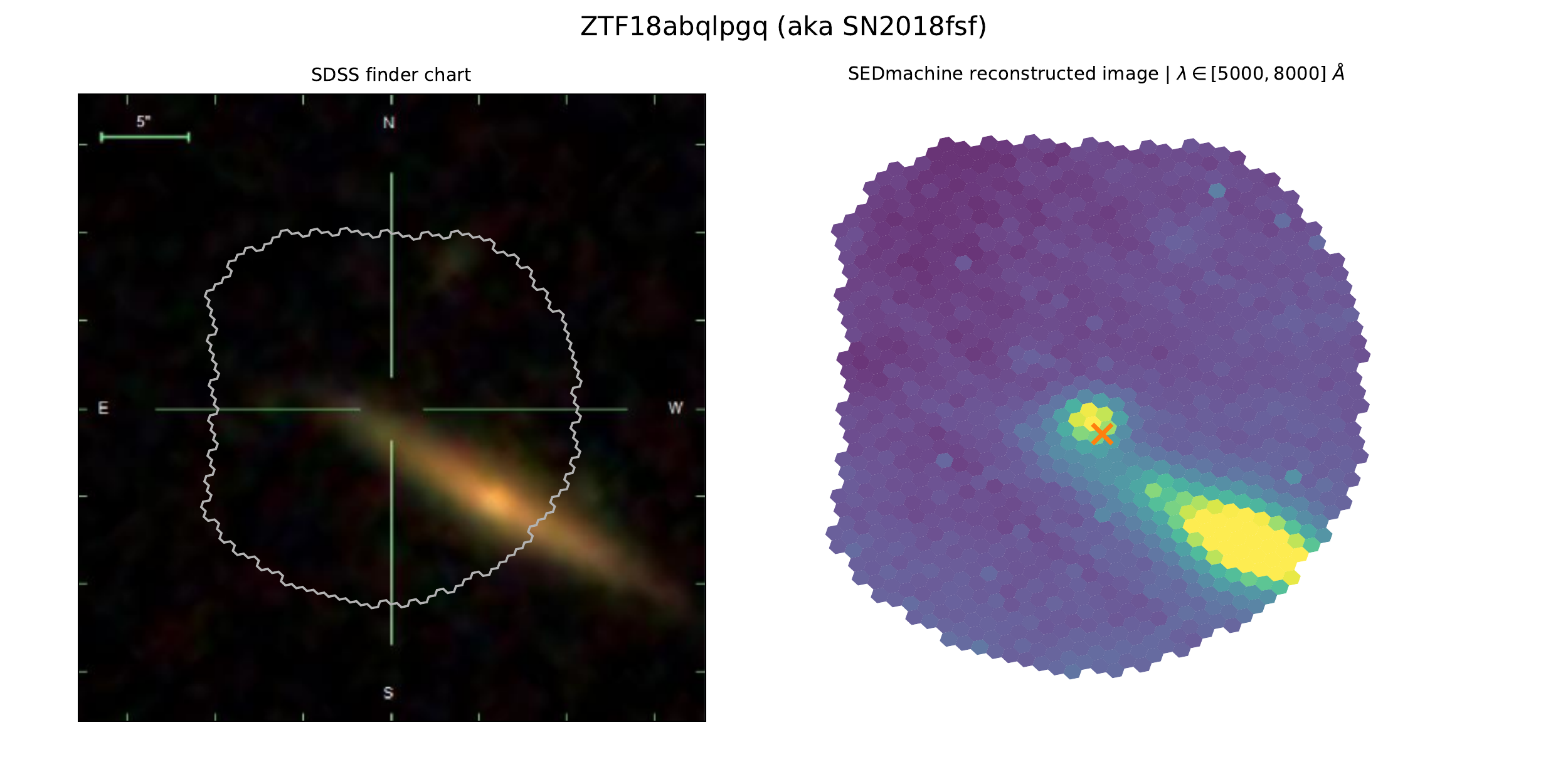}
  \caption{Illustration of the SEDM field of view ;
    ZTF18abqlpgq. \emph{Left}: RGB SDSS image centered at the target
    (ZTF18abqlpgq) location. \mr{The grey line represents the SEDM
    footprint for this target. \emph{Right}: Integrated 2D image
    of the corresponding SEDM 3D-cube. The orange cross shows the
    expected target location used as first guess for the automatic PSF
    extraction. We clearly see the point source near its expected
    position as well as the host galaxy.}}
  \label{fig:fchart_ZTF18abqlpgq}
\end{figure*}

Finally, a rotation of 263$^{\circ}$ is applied to the reconstructed
hexagonal structure such that north is up and east is to the left. A
reconstructed image is shown in fig.~\ref{fig:fchart_ZTF18abqlpgq}, along with
its corresponding optical image.  We recover the galaxy visible
in the SDSS finding chart in the MLA. The MLA also contains the transient, here
ZTF18abqlpgq, aka SN 2018fsf, a $z=.095$ Type Ia supernova near
maximum light.

\subsection{CCD to Cube}
\label{sec:ccd_to_cube}
\begin{itemize}
\item Input:
  \begin{itemize}
    \item[$\bullet$] \pipeinput{crr\_target\_file.fits},
   \item[$\bullet$] \pipeinput{TraceMatch.fits},
   \item[$\bullet$] \pipeinput{HexaGrid.fits},
   \item[$\bullet$] \pipeinput{WaveSolution.fits} ;
  \end{itemize}
\item Output: \pipeinput{e3d\_target\_file.fits}.
\end{itemize}

The cube extraction makes use of the night calibration produced during
daytime to extract the 3D cube from a given CCD image. In addition, to
account for instrumental flexure, the cube extraction has two
independent flexure correction steps, one for the trace location and one
for the wavelength solution. The spatial solution is fixed.  The cube
extraction is therefore done as follows, with each step further detailed in
the following subsections for careful readers:

\begin{enumerate}
\item \textit{Trace-position flexure correction}:
  vertically move together the trace locations to optimize the trace alignment.
\item  \textit{CCD background removal}: mask out regions contained within the
  flexure-corrected traces and construct a background CCD-image based
  solely on the out-of-trace pixels. This background is stored and removed from
  the CCD image.
\item \textit{Cube extraction}: For each flexure-corrected trace, extract its
  corresponding pixel-spectrum and convert it into a wavelength-spectrum
  using the night wavelength solution. Once all the spaxels are
  created this way, combine them into a cube using the nightly spatial
  solution.
\item \textit{Wavelength flexure correction}: Measure the effective
  wavelength position of sky and telluric lines and convert the
  observed wavelength shift $\Delta \lambda$ into a
  horizontal $\Delta i$ pixel shift.
\item \textit{Flexure corrected cube extraction}: Repeat step 3, applying
  the $\Delta i$ pixel shift identified at step 4 when converting pixel
  into wavelength.
\item \textit{Flat field the Cube}: correct the relative spaxel
  responses using the cube-flat made from the dome exposure.
\end{enumerate}

Once done, the cube is stored as a fits file. The file structure is a
simplified version of the flexible euro3d format \citep{euro3d}.  We invite the reader to
use the \pkg{pyifu} python library to open these cubes. A cube fits file
contains five entries:

The primary image contains an $M\times N$ array, where $M$ is the number of
spaxels and $N$ is the length of each spectrum. The corresponding wavelength
array can be reconstructed from the header, using the starting wavelength
and the wavelength step recorded in the primary header as CDELT1 and
CRVAL1, respectively.

The second image, ``VARIANCE'', contains an $M\times N$ data array
corresponding to the spaxel spectral variance.

The third image, ``MAPPING'', has a $2\times M$ data array. This records
the \{x, y\} coordinates of the $M$ spaxel centers.

The fourth image, ``SPAX\_ID'', has an $M$-length array where the spaxel ``id''
is stored. This ``id'' corresponds to the ``trace-id'' and enables the
user to easily map a given spaxel from a 3D-cube to its corresponding
dispersion trace on the CCD 2D-image.

The fifth image, ``SPAX\_VERT'', contains the spaxel vertices.

As in euro3d, the format adopted in \pkg{pyifu} enables
work with cubes of any structure: spaxel positions in the focal
plane are recorded~-- and having spaxels of any shape
--~their vertices are stored. However, our simplified version of
euro3d enforces all spaxels to share the same shape and their spectra
to share the same wavelength binning.

\subsubsection{Trace Position Flexure Correction}

The spaxel trace positions are predefined for the entire night based on
dome observations (see section~\ref{sec:nightcalibration_tracematch}).
However, while the telescope is moving and pointing toward a science
target, it is expected that the instrument alignment will change slightly.
Therefore, the original trace position might not be optimal. Given the SEDM
configuration, vertical misalignment will translate into signal-to-noise
reduction\mr{, as a} fraction of the light will not be included inside
the trace~-- and CCD-background bias (see next subsection) --~ where we
assume that light outside the traces are pure background. On the other
hand, horizontal flexure will cause wavelength miscalibration. While this
might not be optimal, we have decided in \pysedm{} to treat both flexure
effects independently. See section~\ref{sec:ccdtocube_iflexure} for
correction of the horizontal displacement.

\begin{figure}
  \centering
  \includegraphics[width=\linewidth]{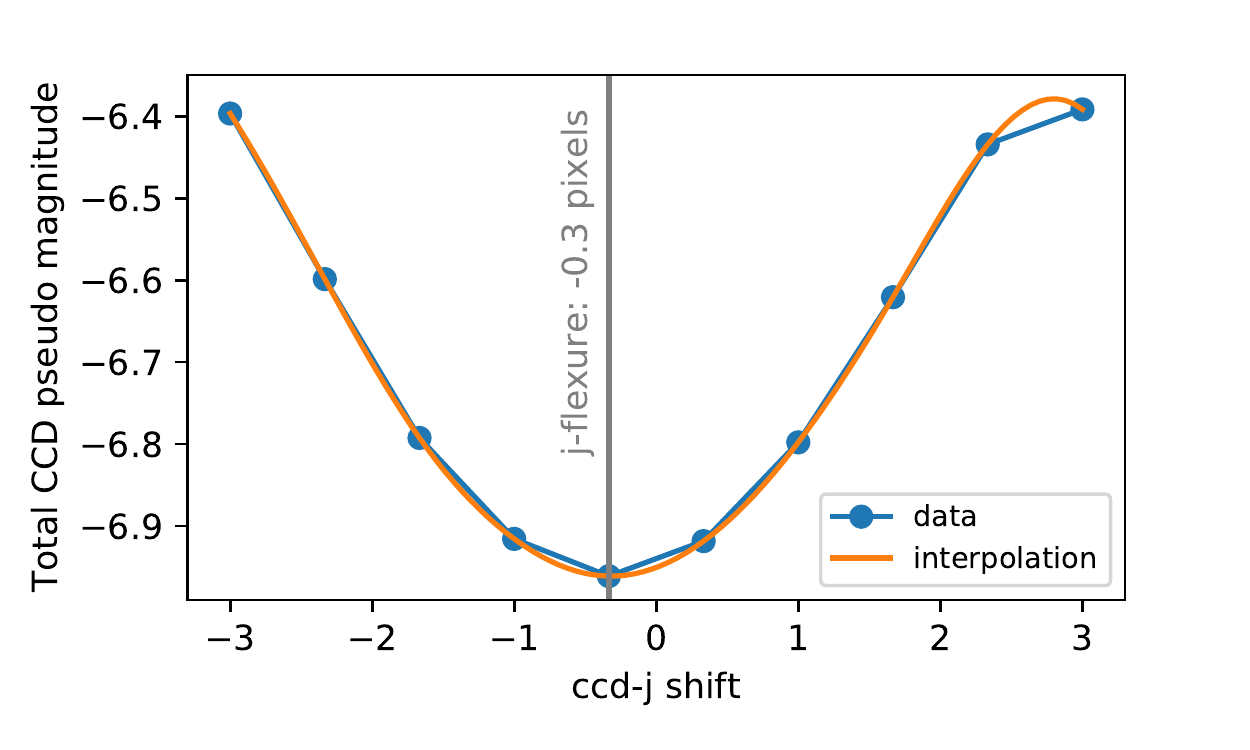}
  \caption{Illustration of the vertical trace flexure procedure;
    here a science observation of ZTF18abqlpgq. Blue markers show the
    pseudo-magnitude computed inside the spaxel traces, the lower the
    brighter. The orange curve is a cubic spline interpolator. The
    grey vertical bar shows the minimum of the cubic interpolation. In
    this illustrative case, the dome-based trace mask must be shifted
    by $-0.3$ pixels to account for vertical flexure.}
  \label{fig:jflexure_ZTF18abqlpgq}
\end{figure}

To measure the vertical trace displacement for a given exposure, we measure
the total flux $T_f$ contained within all spaxel traces while vertically
moving them from -3 to +3 pixels around the default position in 10 regular
steps. Each time, a pseudo-magnitude is defined as $-2.5 \times \log(T_f)$.
For an illustrative observation of
ZTF18abqlpgq, figure~\ref{fig:jflexure_ZTF18abqlpgq} shows the evolution of this
pseudo-magnitude as a function of the vertical shift applied. We interpolate the
10 measured pseudo-magnitudes with a cubic spline to find the vertical
shift minimizing the pseudo-magnitude. In this example the shift is
$-0.3$ pixels; shifts rarely go beyond $\pm 1$ pixel.

In the rest of the cube extraction procedure, the nightly TraceMask
solution will be vertically shifted by $+0.3$ pixels to counterbalance the
vertical instrument flexure.

\subsubsection{CCD Background Subtraction}

After the trace flexure correction, light outside the traces may be
considered to be diffuse. In order to remove this background light from the
actual spaxel signal, we build a CCD-background that will be removed from the
CCD image. This background is constructed by
fitting every 10 column a 
fifth-order Legendre polynomial on pure-background pixels, i.e. those in none of
the spaxel traces. The actual background CCD-image is built by
concatenating these fitting polynomials and interpolating between them with
a horizontal Gaussian kernel with a width of 5 pixels.

\subsubsection{Cube extraction}

For each spaxel, we first create a pixel-spectrum (see definition in
section~\ref{sec:nightcalibration_tracematch}) using the
flexure-corrected 
spaxel trace mask on the background-subtracted CCD-image. We convert pixels
to wavelengths using the wavelength solution to create the
wavelength-spectrum of this spaxel. This wavelength solution is the nightly
wavelength solution onto which we can apply a pixel shift to account for
wavelength flexure (see section~\ref{sec:ccdtocube_iflexure}).

Once all spaxels have their wavelength-spectrum, we merge them into a unique
cube using the nightly HexaGrid for their \{x, y\} position into the focal
plane; x increasing northwards and y increasing westwards.

\subsubsection{Wavelength Flexure Correction}
\label{sec:ccdtocube_iflexure}

Once a cube has been created, we can test and correct the wavelength
solution accuracy by comparing the measured sky and telluric line positions
with their expected locations.

We first randomly arrange the 600 faintest spaxels ($\approx$ half of
the MLA) into 40 spectra averaging 15 spaxels. We assume the sky to be
constant over the IFU given its small field of view. The faintest
spectra are thus the ones for which the sky ratio is the
largest. The random spaxel mixing enables us to average out any
potential spatial structure. This way all 40 spectra are comparable.

\begin{figure}
  \centering
  \includegraphics[width=\linewidth]{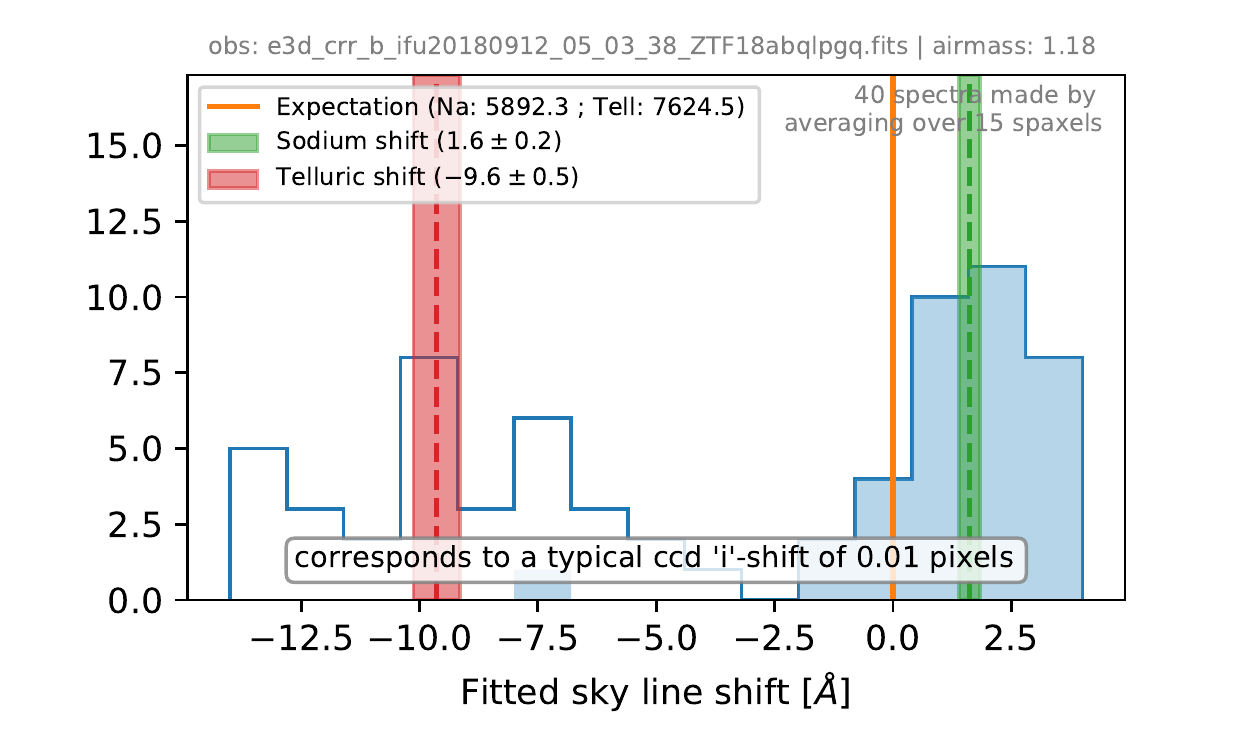}
  \caption{Illustration of the horizontal wavelength flexure procedure;
    here a science observation of ZTF18abqlpgq. 
    Histograms show the measured wavelength
    shift between expected and measured sky sodium lines (blue) and
    telluric absorption (white). The measurements are made on the 600
    faintest spaxels ($\sim$half of the MLA) into 40 spectra averaging 15
    spaxels. The orange vertical bar shows the average shift in
    wavelength. The text indicates the conversion in terms of
    horizontal pixels. In the illustrative case there is barely any
    wavelength flexure correction, which rarely exceeds a fraction of
    a pixel.}
  \label{fig:iflexure_ZTF18abqlpgq}
\end{figure}

We fit these 40 spectra with the sum of a fifth-degree polynomial and two
Gaussian lines: one in emission for the strong sky sodium line, and
one in absorption, for the main O$_{2}$ telluric complex. The centroid of
these lines is allowed to shift. We show in
fig.~\ref{fig:iflexure_ZTF18abqlpgq} the difference between the
expected measured lines and their positions, assumed to be 5892.3 \AA\
and 7624.5 \AA\ for the sodium and the telluric lines,
respectively. We consider the effective wavelength shift to be the
average shifts, weighted by the square of the inverse of the
mean shift error. The wavelength shift, $\Delta \lambda$, is then
converted into a horizontal pixel shift, $\Delta i$. In the example
shown in fig.~\ref{fig:iflexure_ZTF18abqlpgq}, this shift is as small as
$+0.01$ pixels. Typical $\Delta i$ shifts are $\approx 0.5$ pixels,
and are rarely above 1 pixel.

We note in fig.~\ref{fig:iflexure_ZTF18abqlpgq} that both line shifts are not
identical. While this is not always the case, it actually is expected. The
O$_{2}$ telluric absorption is a complex combination of numerous blended lines
whose relative amplitudes depend on atmospheric conditions (see
section~\ref{sec:fluxcal}). Therefore the exact telluric line
centroid varies, which makes it a questionable reference candidate. We
nonetheless decided to use it for two reasons: First and most importantly,
the telluric
O$_{2}$ complex is always clearly detected, even during short exposures and for
cases in which the sky brightness is small in comparison to the targeted
source --~including its host environment. Consequently, keeping in mind that
\pysedm\ is designed to be a fully automated pipeline and given the low
wavelength resolution of the SEDM, the reliability of the detection of the
telluric line largely counterbalances any inaccuracy of a few angstroms.
Second, because we are applying a single global pixel shift as a wavelength
flexure correction, using the telluric line along with the sodium sky line
enables us to have two wavelengths to anchor the wavelength shift.
We note that part of the discrepancy observed in
fig.~\ref{fig:iflexure_ZTF18abqlpgq} could also be due to the
wavelength flexure correction being more complex than the
simple pixel shift assumed in \pysedm. However, given the low
wavelength resolution of SEDM, this instrument should not be used for accurate
redshifts or element-velocity analyses and we therefore decided to keep the
simple and reliable single shift technique.

\section{Spectral Extraction from 3D Cube}
\label{sec:extract_star}

Within two minutes of the end of a science exposure, a 3D cube
fits file is created following the \pysedm\ cube extraction detailed
in section~\ref{sec:cube_extraction}.

Each spaxel of the 3D data cube includes three intensity components, each one
characterized by its own spatial signature: (1) the target flux,
from a point source located close to the center of the MLA,
(2) the night sky flux, a spatially flat component over the entire
IFU field of view, and (3) the (potential) host galaxy flux, a
structured background. The core purpose of \pysedm\ is to
automatically extract the point source spectrum to enable the
typing of the transient with no human interaction. However, for full
flexibility and reliability, the code allows human interaction
for potential fine-tuned re-extractions
of complex cases (see section~\ref{sec:humain_interaction}).

We present in the following subsections the transient
spectrum-extraction process. We start section~\ref{sec:guider} by showing how
we make use of the imaging channel of the SEDM to estimate the target location
in the MLA. Then, in section~\ref{sec:3dpsf} we describe the 3D point
spread function (3D-PSF) modeling and the actual source extraction.
The flux calibration procedure is described in
section~\ref{sec:fluxcal}, and in section~\ref{sec:humain_interaction} we
describe the manual operation of the transient extraction.

The \pysedm\ point source extraction pipeline is largely inspired by
the Nearby Supernova Factory \citep{aldering_2002}
collaboration development of the  SuperNova Integral Field Instrument
\citep[SNIFS,][]{lantz_2004}; see also the detailed work presented
(in French) in \citet{copin_hdr}.
We have built a simplified version of their 3D-PSF extraction (see
section~\ref{sec:psf_profile}). We have also adapted their
telluric absorption correction method
\citep[see section~\ref{sec:fluxcal}, and][]{buton_2013}, and we are
using their atmospheric differential refraction correction algorithm
\citep[see section~\ref{sec:adr}, and][]{copin_hdr}.

\subsection{Target location in the Multi-Lens Array}
\label{sec:guider}
\begin{itemize}
\item Input: [ \pipeinput{guider\_files\_of\_crr\_target.fits} ];
\item Output: \pipeinput{metaguider\_astrometry\_crr\_target.fits}.
\end{itemize}

The P60 telescope pointing is not accurate enough to perfectly place
the target in a pre-defined location on the MLA.
Furthermore, we can not directly solve the astrometric
solution of the MLA, as it rarely contains more than the target and its
potential host in the $32\times32\,\mathrm{arcsec}^2$ field of
view. It is therefore non-trivial to know
where the target is located without inspecting the data cube --~which
would make the pipeline non-automatic.

The simplest cases are those in which the target is the brightest or the
sole visible object in the MLA. Observations of standard stars are a typical
example. For them, one can simply take the brightest spaxel as a good
initial guess of the target location at a reference wavelength (see
section~\ref{sec:adr}). The spaxel brightness is
defined as the integrated flux between two wavelength boundaries
--~typically $5000$ to $8000\,\AA$ for \pysedm.
In practice, to ensure the  robustness of the automatic pipeline, we
do not use the brightest spaxel position but the median position of
the five brightest ones. This avoids single-spaxel issues, such as cosmic ray contamination. This simple target positioning
is referred to as the ``brightness position'' method.

General cases are the most complex. A typical target is accompanied
by and possibly overlapped by a host galaxy detectable in the MLA.  In 
many cases the targeted transient is not the brightest object in the MLA.
For such cases, we need
external information to automatically locate the target
position. Fortunately, the SEDM has a parallel imaging channel.
This channel is used for guiding
during IFU exposures. As detailed in \cite{sedm_paper}, this
photometric channel, called the rainbow camera (RCAM), has a
$13\times13\,\mathrm{arcmin}^{2}$ field of view split into four
quadrants, each having a different optical filter ($u,\ g,\ r,\ i$) and
having the IFU pick-off mirror mounted in the center.
Thus, the RCAM is illuminated from the P60 focal plane simultaneously with
the $32\times32\,\mathrm{arcsec}^{2}$ field of view of the IFU.
The position of the IFU pick-off \mr{prism} is fixed, allowing one to
project a WCS solution calculated from the guider images onto the MLA.
This is what is done in \pysedm.

At the end of each science
acquisition, we create a median stack of all guider images (one taken every 30s), and we measure the astrometric solution of this ``meta-guider'' image
using a local installation of the astrometry.net software \citep{astrometrynet_paper}.  Since we know the target
coordinates, we are able to predict the RCAM CCD pixel coordinates
$\mathrm{\{x_{rc},\,y_{rc} \}}$ where the target should be located. Note
that, if the telescope pointing is correct, the target will not be visible in the
RCAM image since its light has been reflected by the pick-off mirror onto
the MLA.
Thus, we use the following transformation to project RCAM pixel coordinates
$\mathrm{\{x_{rc},\,y_{rc} \}}$ onto MLA cube
coordinates at a reference wavelength of $7000\,\AA$ $\mathrm{\{x,\,y\}}$:
\begin{equation}
\begin{bmatrix}
\mathrm{x}\\
\mathrm{y}
\end{bmatrix}
=
\begin{bmatrix}
p_1 & p_2\\
p_3 & p_4
\end{bmatrix}
\begin{bmatrix}
\mathrm{x_{rc}} - x_{0}\\
\mathrm{y_{rc}} - y_{0}
\end{bmatrix}
\end{equation}
where $p_i$, the transformation parameters, and $\{x_0,\,y_0\}$, the
centroid shift coordinates, have been derived based on observations of
tens of standard stars for which $\mathrm{\{x,\,y\}}$ coordinates have
been measured using the ``brightness position'' method. The IFU pick-off
mirror is only 5 mm square and is rigidly mounted to the central support
of the RCAM filter assembly, so we expect no significant flexure
in its position relative to the RCAM images.  Therefore these six
parameters are fixed but nonetheless must be updated if any part of the
optical system (including the CCDs) is adjusted.

With this method, referred to as the ``guider-astrometry position'' method, we are
able to predict the position of the transient on the MLA to within a couple
of arc seconds.  This is good enough to avoid confusion
when automatically extracting a point source within its environment
(see section~\ref{sec:psf_profile}). The ``guider-astrometry position'' method is
illustrated in fig.~\ref{fig:fchart_ZTF18abqlpgq}, where we see that the
expected point source position, \mr{marked as an ``x''}, appears about
1 spaxel ($\approx  0.75$ 
arcsec) southwest from the true PSF centroid.

To ensure the \pysedm\ pipeline robustness, if the ``guider-astrometry
position''
method  fails for any reason --~could not access the guider images,
astrometric solution of the meta-guider did not converge, etc~--
we then use the ``brightness position'' method as a backup
solution. In such cases --~less then 5\% of the time~-- a warning
is placed in the header as well as on the validation plots (see
section~\ref{sec:quality_control}). These cases usually require manual
re-extraction 
if the transient is faint (see section~\ref{sec:humain_interaction}).
On rarer occasions, the P60 acquisition offset fails, and we find with the
``guider-astrometry position'' method that the target is outside the MLA.
In this case a warning is raised, the target spectral extraction outputs
a null spectrum, and \pysedm\ reports the target as ``failed''. Most
of the time, this is caused by poor weather condition.

\subsection{3D PSF extraction}
\label{sec:3dpsf}
\begin{itemize}
\item Input:
  \begin{itemize}
    \item[$\bullet$] \pipeinput{e3d\_target\_file.fits},
   \item[$\bullet$] ``source position'' (automatic or manual)
  \end{itemize}
\item Output:
    \begin{itemize}
      \item[$\bullet$] \pipeinput{spec\_target\_file.fits} ;
      \item[$\bullet$] \pipeinput{e3d\_psfmodel\_file.fits}.
      \end{itemize}
\end{itemize}

The 2D PSF profile modelling is presented in
section~\ref{sec:psf_profile}. We then detail in
section~\ref{sec:psf_background} how this profile is actually matched
to the data accounting for PSF ellipticity and target background.
In section~\ref{sec:adr}, we explain how we model the evolution of the
PSF centroid as a function of wavelength. The complete spectral
extraction procedure is detailed in
section~\ref{sec:psf_extraction}.


\subsubsection{PSF Profile}
\label{sec:psf_profile}

Given an IFU wavelength-slice or meta-slice, and calling $r$ the
elliptical radial distance from the PSF center, we
parameterize our radial PSF function $\mathcal{P}$ as a linear combination of
a Gaussian and a Moffat distribution:
\begin{equation}
\label{eq:psf}
\mathcal{P}(r; \alpha, \sigma, \rho_{nm}) =
\frac{\rho_{nm}}{1+\rho_{nm}} \times \mathcal{N}(r ; \sigma) +
\frac{1}{1+\rho_{nm}} \times \mathcal{M}(r ; \alpha, \beta(\alpha))
\end{equation}
Where $\mathcal{N}(r, \sigma)$ is the normalized Gaussian distribution
center of 0 with a scale of $\sigma$ --~from scipy.stats.norm~-- and
$\mathcal{M}(r, \alpha, \beta(\alpha))$ is the normalized Moffat
distribution expressed below; $\rho_{nm}$ is the relative amplitude of
the two distributions:
\begin{equation}
\label{eq:moffat}
\mathcal{M}(r ; \alpha, \beta) = 2\,\frac{\beta-1}{\alpha^2} \left[ 1 +
  \left(\frac{r}{\alpha} \right) ^2 \right]^{-\beta}
\end{equation}
An illustration of the ``Gaussian+Moffat'' PSF model is presented in
fig.~\ref{fig:psfprofile_ZTF18abqlpgq}; see details on the complete
model in section~\ref{sec:psf_background}.

To limit the number of correlated parameters in
the PSF determination, the Moffat $\beta$ parameter in eq.~\ref{eq:psf}
is parameterized as a linear function of $\alpha$, such that $\beta =
b_0+\alpha\times b_1$ as inspired by \cite{buton_2013}. $b_0$ and $b_1$ have been determined from high
signal-to-noise standard star observations to be $0.25$ and  $0.63$,
respectively.  These values are in agreement with the corresponding
parametrization for SNIFS presented in \cite{buton_2013}.

\subsubsection{PSF+Background slice model}
\label{sec:psf_background}
Our primary aim is rapid typing of the observed transient with the SEDM.  Thus, we
are not taking final reference images, i.e., observation of the host galaxy
long after the transient has faded away. Therefore, we cannot apply the
host removal technique presented in \cite{bongard_2011}.

Instead, to account for the spatially flat --~sky~-- and
structured --~host~-- background signals associated with the target
point source, we fit a 2D multi-degree
polynomial simultaneously with the PSF profile, as done in \cite{bailey_2009}.
Three levels of complexity have been implemented: a
simple flat continuum, a tilted-plane, and a curved-plane, which have
one, three and five free parameters, respectively. The default background in \pysedm\ is the
tilted plane \citep{bailey_2009} and will be the one used in this paper if not
specified otherwise.

Altogether, the ``Gaussian/Moffat' + Tilted Plane'' 2D model of a cube
wavelength-slice (or meta-slice) has 11 ($\pm 2$) free parameters:
\begin{itemize}
 \item 2 for the PSF centroid position $\{x_{psf}^0, y_{psf}^0\}$ ;
 \item 2 for $e$ and $\theta$, the PSF ellipticity and ellipse
   rotation parameter, respectively ;
  \item 1 for the PSF profile amplitude (the flux);
 \item 3 for the normalized PSF radial profile (see
   section~\ref{sec:psf_profile}) ;
 \item 3 for the background (1 if flat continuum, 5 if curved plane).
\end{itemize}

Using this model, we can create a 2D slice model by assuming that each
slice ``spaxel'' has a flux value corresponding to that of its center
in \{x, y\}  coordinates. For the radial parametrization of the PSF, we
convert \{x, y\} to the centroid elliptical coordinate defined by the
radial distance $r=\sqrt{x_{ell}^2 +
  y_{ell}^2}$ which gives the following transformation to $x_{ell}$ and $y_{ell}$:
\begin{equation}
\begin{bmatrix}
\mathrm{x_{ell}}\\
\mathrm{y_{ell}}
\end{bmatrix}
=
\begin{bmatrix}
\cos(\theta) & \sin(\theta)\\
-\sin(\theta) &\cos(\theta)
\end{bmatrix}
\begin{bmatrix}
\mathrm{x} - x_{psf}^{0}\\
\mathrm{y} - y_{psf}^{0}
\end{bmatrix}
\end{equation}

\begin{figure}
  \centering
  \includegraphics[width=\linewidth]{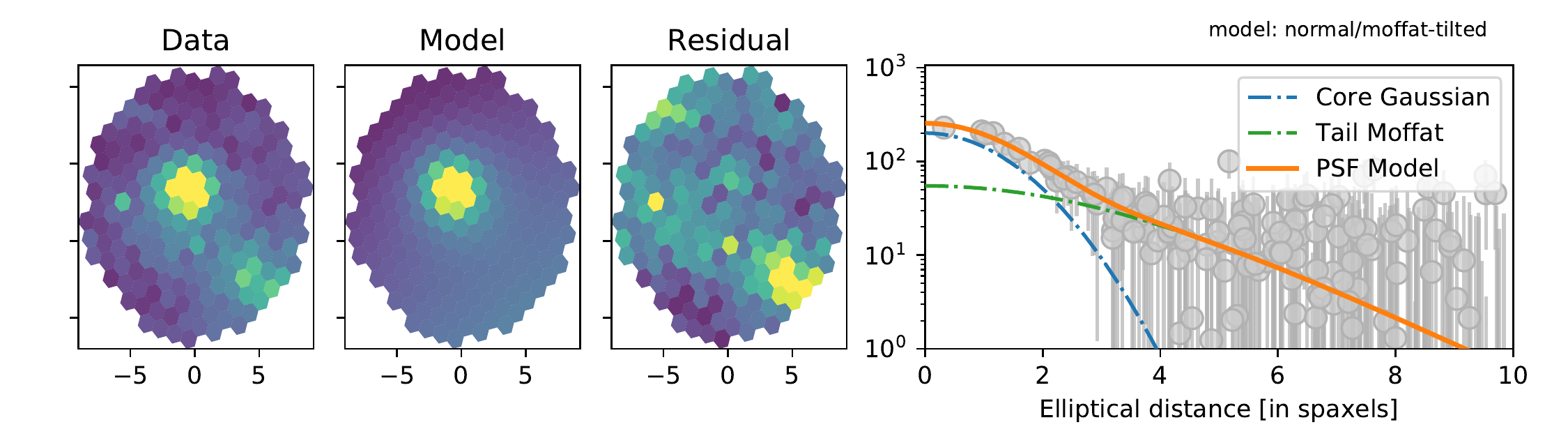}
  \caption{Illustration of the PSF model used to extract to target
    point source from the cubes 3D, here showing the slice [5000, 5500] $\AA$
    of the part of ZTF18abqlpgq's $x,y,\lambda$-cube used for PSF
    fitting (see text). 
    \emph{From left to right}
    data slice, model slice (PSF+tilted-plane background), residual
    slice and PSF profile. The PSF profile axes on the right hand side
    present the background-subtracted spaxel flux (grey marker with
    error bars) [in pseudo-ADU, see text] as a function of the 
    elliptical distance to the best fitted center of the PSF. 
    The orange curve is the best fitted PSF profile, decomposed into
    Moffat (green) and Gaussian (blue).
    The small structure visible on the lower right of the residual
    slice panel is the host galaxy signal (see fig.~\ref{fig:fchart_ZTF18abqlpgq}).}
  \label{fig:psfprofile_ZTF18abqlpgq}
\end{figure}

An illustration of the ``PSF+Tilted Plane'' model is presented
in fig.~\ref{fig:psfprofile_ZTF18abqlpgq}. We see on the right panel of the
figure that the ``Gaussian+Moffat'' profile well describes the data: the
Gaussian profile dominates the core of distribution while the Moffat
profile well matches the PSF tail. We note that a good
PSF radial match also implies a good fit of both the PSF centroid
$\{x_{psf}^0, y_{psf}^0\}$ and of the PSF elliptical parameters $e$
and $\theta$.

The ``Model'' panel represents the model-slice reconstruction where
one can distinguish the 2D reconstruction of the PSF on top of the
tilted plane.  Note the south-west / north-east inclination of the
background. The goodness of fit is demonstrated in the
``Residual'' panel where we see no structure left at the PSF
location. The bright signal in the lower-right --~also visible in the
``Data'' panel~-- is actually a part of the host galaxy (see
fig.~\ref{fig:fchart_ZTF18abqlpgq}).
As further discussed in section~\ref{sec:psf_extraction}, only the part
of the cube near the target is fitted.

\subsubsection{ADR correction}
\label{sec:adr}

The atmosphere acts as a prism and bends light by an amount that depends
on wavelength. The
image is then shifted from its original position $\mathrm{\{x_{ref},
  y_{ref}\}}$ toward the zenith to a degree dependent on the
 wavelength, the
atmosphere refraction index $n$, and the target airmass
$z$. To first order, the mean atmospheric refraction is
accounted for by the telescope pointing model. Hence, only the atmospheric
differential refraction (ADR), related to wavelength variations of the refraction
effect, has to be taken into account during the spectral extraction
from the 3D cube: see details in \cite{copin_hdr}.

Theoretically, three variations affect the observations: (1) temporal
variations ~--the evolution of $n$ and  $z$ as a function of time
during the exposure, (2) spatial variations ~--the variations of $n$ and
$z$ within the field of view, and (3) chromatic variations ~--
the wavelength dependence of $n$. In the SEDM, the spatial variations are
negligible. With less than 1 h exposures, the temporal variation will
induce a slight blurring that will be, to first order, accounted for
by the PSF model. Consequently, only the chromatic variation has to
be modeled.

The effect of the chromatic ADR has been extensively studied
\citep[e.g.,][]{filippenko_1982}, including specifically for 3D spectroscopy
\citep[e.g.,][]{arribas_1999}. The target position in the MLA as a
function of wavelength $\{x(\lambda),\,y(\lambda)\}$ is thus given by:
\begin{align}
\label{eq:adr}
x(\lambda) =&  x_{ref} - \frac{1}{2}\left( \frac{1}{n^2(\lambda)} -
              \frac{1}{n^2(\lambda_{ref})} \right)  \times
        \tan(d_z) \sin(\theta),\\
y(\lambda) =&  y_{ref} - \frac{1}{2}\left( \frac{1}{n^2(\lambda)} - \frac{1}{n^2(\lambda_{ref})}\right)  \times
        \tan(d_z) \cos(\theta),
\end{align}
where $d_z=\arccos(z^{-1})$ is the zenith distance, $\theta$
is the parallactic angle, and $\lambda_{ref}$ is a reference wavelength
set at $7000\,\AA$ in \pysedm. Hence, $\mathrm{\{x_{ref},
  y_{ref}\}}$ is the target position in the MLA at $\lambda_{ref}$. We
use the air refraction index $n(\lambda)$
modeled by \cite{stone_adr}\footnote{NIST:
  https://emtoolbox.nist.gov/Wavelength/Documentation.asp},
which depends on pressure, temperature and relative humidity. These
atmospheric parameters are provided in the FITS image header.

Deriving the $\{x(\lambda),\,y(\lambda)\}$ consequently requires 4
free parameters: the reference position $\mathrm{\{x_{ref}}$ and
$\mathrm{y_{ref}\}}$, the effective airmass, $z$ of the exposure, and
the parallactic angle, $\theta$. The ADR fitting procedure is detailed
in section~\ref{sec:psf_extraction}.

\subsubsection{Point Source Spectral Extraction}
\label{sec:psf_extraction}

The point source extraction is made in three steps:
(1) fit the ``PSF+Background'' model on several large wavelength meta-slices,
(2) model the run of the PSF parameters with wavelength, and (3) fit the PSF
amplitude at each wavelength slice of the 3D cube.
The derived PSF amplitude as a function of wavelength is equivalent to the
target spectrum.

For step (1), we split the 3D cube into six meta-slices with
regular wavelength bins between $4500\,\AA$ and $7000\,\AA$. This
wavelength range is where the SEDM's efficiency
is at its maximum \citep{sedm_paper}, yet extends blue
enough to well constrain the ADR parameters. We note that the number of 
meta-slices and the wavelength range used can be manually provided by
the user in the case of a non-automatic extraction.
The rightmost panel of fig.~\ref{fig:psfprofile_ZTF18abqlpgq} shows the
PSF profile fit
of the second bluest meta-slice ($4916, 5333\ \AA$) made during
the automatic extraction of ZTF18abqlpgq's spectrum.

\begin{figure}
  \centering
  \includegraphics[width=\linewidth]{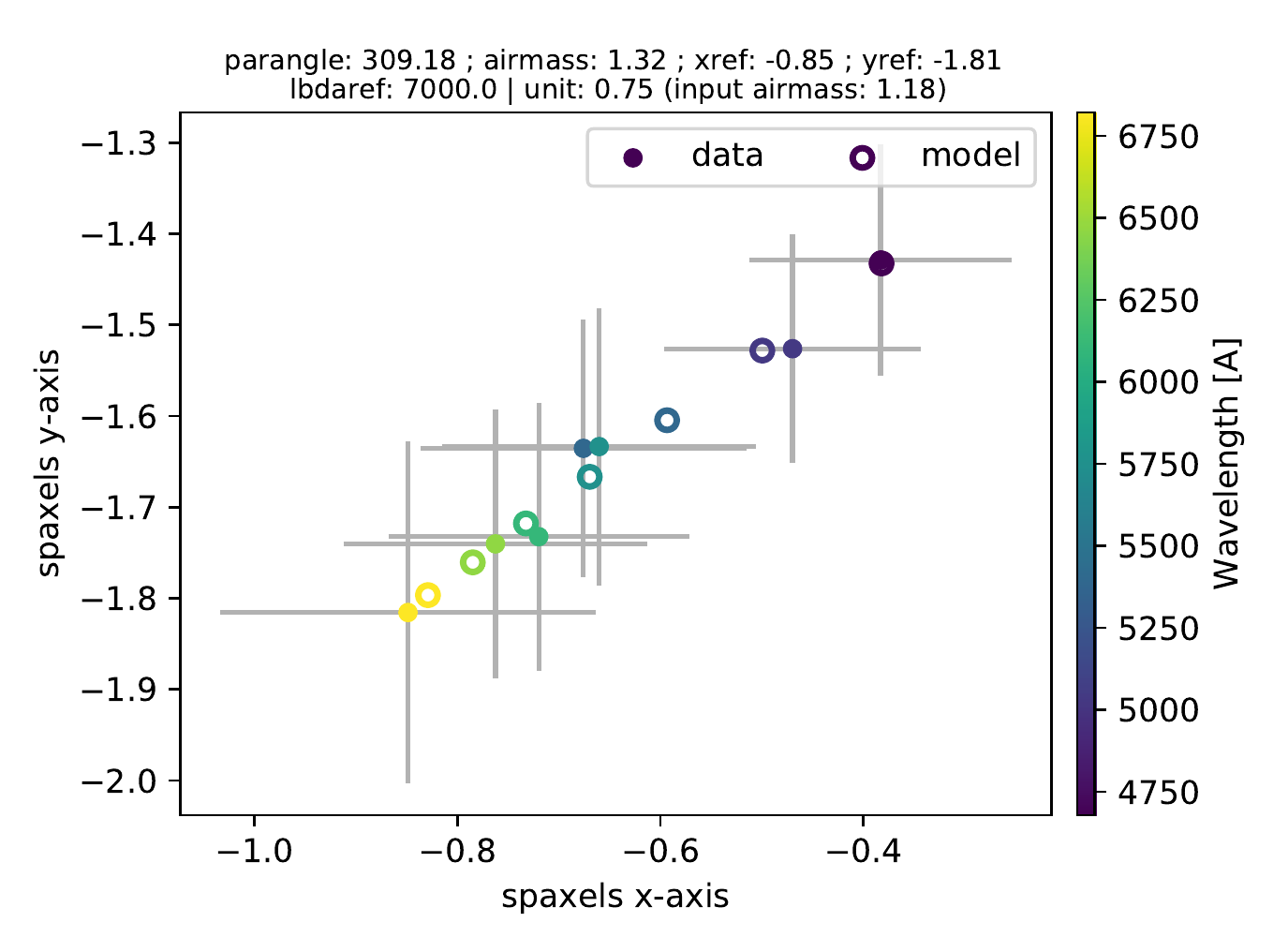}
  \caption{Best fitted PSF centroid positions (full-markers) estimated
    on 6 meta-slices of ZTF18abqlpgq's $x,y,\lambda$-cube during the
    first stage of the automated PSF extraction. Marker colors
    indicate the effective wavelength of the slice. Atmospheric
    differential refraction causes the true position of the target in
    the MLA to vary with \mr{wavelength}. Our best model of this evolution is
    shown as open markers. This automated ADR parameter estimation
    allows us to predict the target centroid to an accuracy of a few percent of a spaxel.}
  \label{fig:adr_ZTF18abqlpgq}
\end{figure}

For step (2), we fit each of the PSF
parameters as a function of wavelength using the meta-slices from the first step.
This contains two parts: the ADR component and the PSF profile
component. For the ADR component, we use
the PSF centroids $\{x_{psf}^\lambda, y_{psf}^\lambda\}$
measured per meta-slice (made during step (1)) to fit the ADR free parameters
$\mathrm{\{x_{ref}, y_{ref}\}}$, $z$, and $\theta$ (see section~\ref{sec:adr}). During
the fit, $z$ and $\theta$ are bounded by their corresponding
measurements stored in the header at the beginning and the end
of the exposure.  The reference position, $\mathrm{\{x_{ref},  y_{ref}\}}$,
should be within 3
spaxels of the expected target position (see section~\ref{sec:guider}).
We will use the best fitted ADR parameters in the next step to fix the PSF
centroid position as a function of wavelength.
As illustrated in fig.~\ref{fig:adr_ZTF18abqlpgq} we are able to
accurately predict the PSF centroid position as a function of
wavelength to better than a tenth of a spaxel.
This is further evidence that the PSF model used is reasonable, as the
physical modeling of the ADR matches the observed centroid evolution.
For the PSF profile component, we use the measured profile values for each
meta slice to model their behavior as a function of wavelength. The
ellipticity parameters $e$ and $\theta$ are assumed to be fixed with
wavelength and are therefore
set to their mean values. To model the dependence of the
remaining three parameters $\alpha$, $\sigma$, and $\rho_{nm}$ on wavelength, we
use standard stars which are bright enough to divide into many meta-slices
and still obtain a good PSF fit.
The Moffat $\alpha$ and the ratio parameter $\rho_{nm}$ turn out to be
largely achromatic. While a linear fit would have been more flexible, we
decided to use the achromatic approximation to increase the \pysedm\
automatic pipeline robustness. The Gaussian $\sigma$ parameter
is evolving as $\sigma(\lambda)\simeq\sigma_{ref}
\left(\lambda/\lambda_{ref}\right)^{-1/5}$ as expected from Kolmogorov
seeing modeling.  This is expected, since the Gaussian traces the core
of the PSF profile \citep{fried_1966, tokovinin_2002}.

In step (3), we perform the actual flux extraction. The
chromaticity of the PSF parameters, including centroids and ellipticity, have
been fixed during step (2).  The sole remaining free parameters are the
background parameters and the PSF intensity. These are fitted
iteratively for each wavelength slice. The PSF amplitude as a
function of wavelength corresponds to the target spectrum in the cube's
native, i.e. non-flux-calibrated, units.

\begin{figure}
  \centering
  \includegraphics[width=\linewidth]{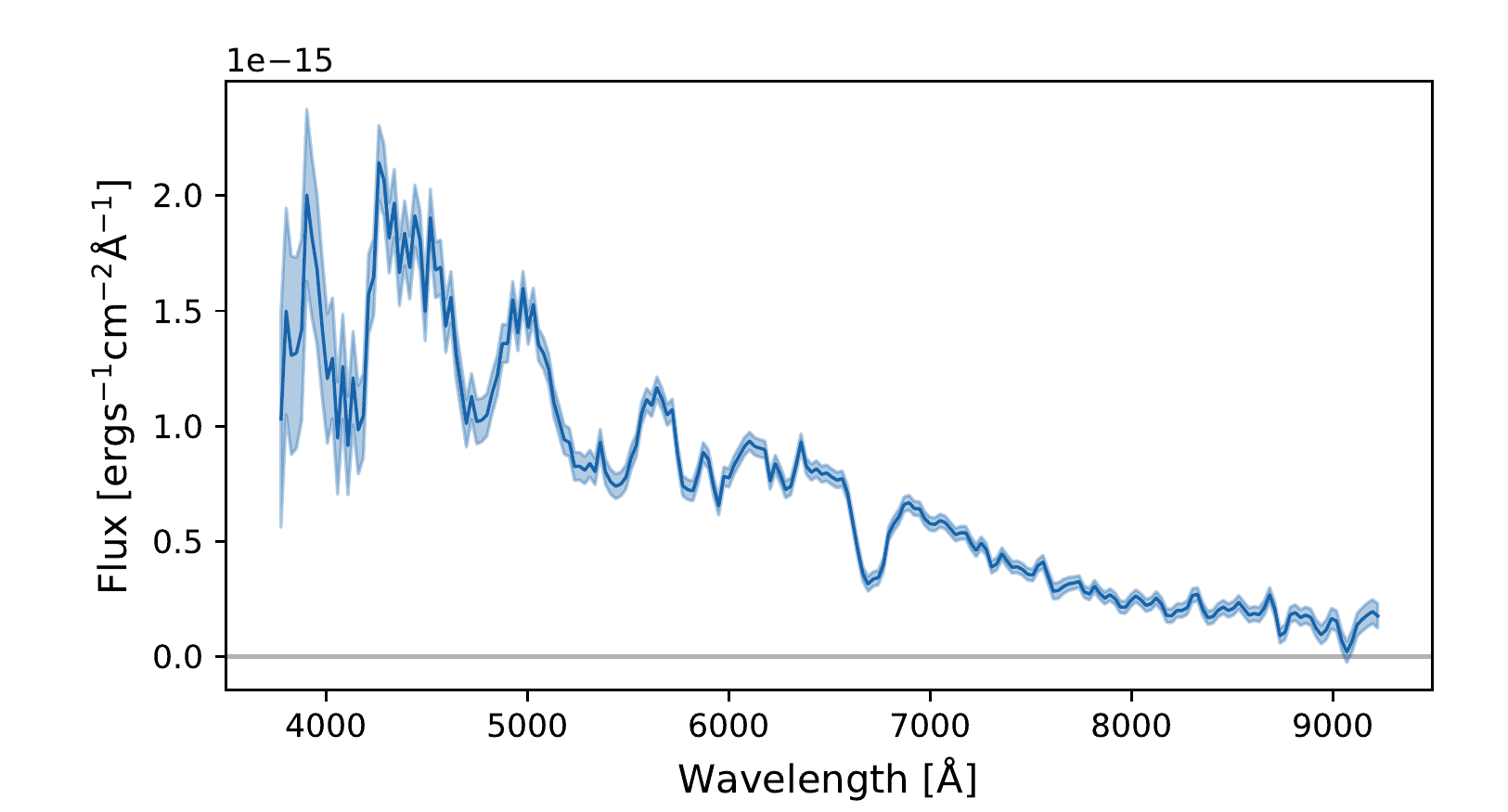}
  \caption{Flux-calibrated spectrum of ZTF18abqlpgq automatically
    extracted using \pysedm{}. The shaded blue band represents the
    flux error.}
  \label{fig:spec_ZTF18abqlpgq}
\end{figure}

The resulting 3D PSF model cube and the target spectrum are saved as
fits files after flux calibration (see section~\ref{sec:fluxcal}). The
extracted flux-calibrated spectrum of  ZTF18abqlpgq is shown
in fig.~\ref{fig:spec_ZTF18abqlpgq}. \mr{This r-band 19~mag spectrum
  is a typical 
good example of what ``SEDM+\pysedm'' can automatically acquire in 
2430s.}

\subsection{Flux Calibration}
\label{sec:fluxcal}
\begin{itemize}
  \item Input files: \pipeinput{calspec\_file};
  \item Input pipeline product: \pipeinput{spec\_target\_file.fits} ;
  \item Created product: \pipeinput{fluxcalibration\_target\_file.fits}.
\end{itemize}

The spectrum extracted at the end of section~\ref{sec:psf_extraction}
are in units we will call ``pseudo'' ADU: ``pseudo'', because several
renormalizations have been applied during the extraction process. The
purpose of the flux calibration step is to recover the target spectrum
in physical units. This means removing the instrumental response
$\mathcal{C}(\lambda, t)$ and the telluric absorption
$\mathcal{T}_{\mathrm{atm}}(\lambda, t, z)$, which depends on airmass
$z$. We developed a much-simplified
version of the method from \cite{buton_2013}, since our goal is focused on
rapid and reliable transient classification rather than accurate
spectrophotometry.

During late twilight, as well as a few times during the night (it
varies depending on the schedule), we take short observations of
standard stars from the calspec catalog \citep{calspec_ref}.
The flux calibration step of \pysedm{} compares calspec-calibrated
spectrophotometry to the observed SEDM spectra to derive the instrumental
response of the SEDM and to solve for the telluric correction for
the standard observation.

During the automatic pipeline processing, each science target is
flux-calibrated using the $\mathcal{C}$ from the telluric-corrected
standard star observation and is also telluric-corrected using
$\mathcal{T}$ based on its own airmass.

Following \cite{buton_2013}, we use the high resolution telluric spectrum from
the Kitt Peak National Observatory\footnote{http://www.noao.edu/kpno/}.
This spectrum is split into two categories: the O$_{2}$ wavelengths, and
the H$_{2}$O wavelengths \citep[see table 1 of ][]{buton_2013}.
Wavelengths not contained in either of the two components are assumed to be free from
telluric absorption. The relative amplitude $c_i$ and the airmass dependence $\rho_i$
are free parameters for each of the two parts (i = O$_{2}$ or H$_{2}$O), such that:
\begin{equation}
  \label{eq:telluric}
  \mathcal{T}(z) = \mathcal{T}_{O_{2}} \times \left(c_{O_{2}} \times z^{\rho_{O_{2}}}) \right) +
    \mathcal{T}_{H_2O} \times \left(c_{H_2O} \times z^{\rho_{H_2O}}) \right),
\end{equation}
where $z$ is the airmass and $\mathcal{T}_i$ is the telluric spectrum for which the
fluxes of wavelengths not contained within the corresponding wavelength-ranges
are set to 0. We model $\mathcal{C}$ as a 20$^\mathrm{th}$ degree Legendre polynomial.

\begin{figure}
  \centering
  \includegraphics[width=\linewidth]{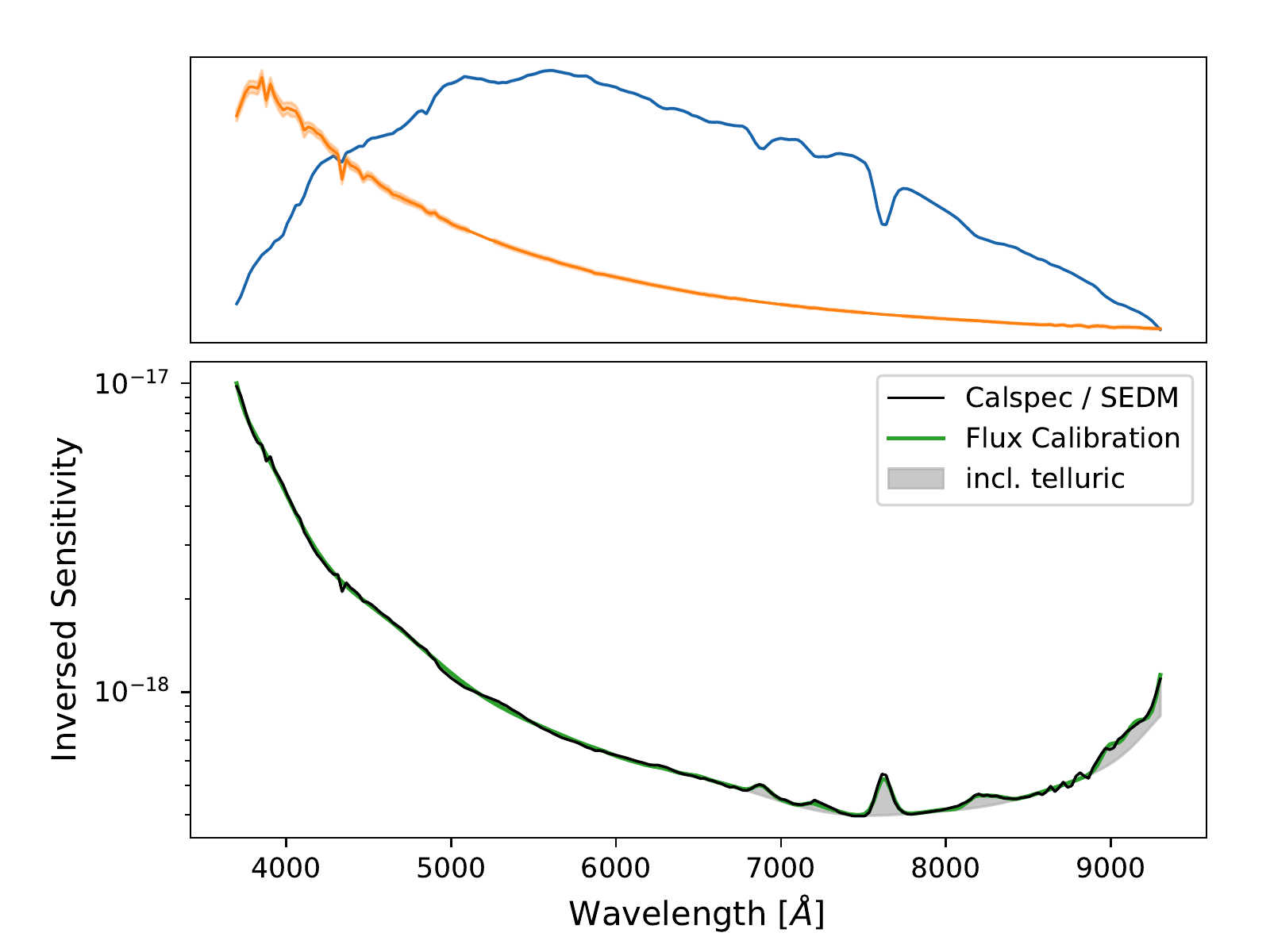}
  \caption{Illustration of the flux calibration procedure. \emph{Top}:
    In blue, extracted standard star spectrum in pseudo-ADU of
    BD+33d2642; in orange, flux-calibrated CalSpec spectrum of the
    same star convolved at the SEDM wavelength resolution used as flux
    reference. \emph{Bottom} In
    black, ratio between CalSpec and SEDM spectra (top). The green
    curve is the full flux calibration model (continuum+telluric). The
    influence of the telluric correction is shown as a grey shaded
    area; hence the lower of the grey area represents the continuum
    component of the flux calibration. When flux-calibrating a science
    target, the intensity of the telluric component (in grey) will vary
    to match that expected at the science target airmass.}
  \label{fig:std_oct_stdstars}
\end{figure}

The estimation of $\mathcal{C}$ and $\mathcal{T}$ parameters is made as follows.
Given a standard star observation $S$ in pseudo-ADU, we load its corresponding calspec 
spectrophotometric spectrum that we convolve to the SEDM resolution $S_{Ref}$.
We then simultaneously fit for the \mr{instrumental response} $\mathcal{C}$ and telluric $\mathcal{T}$ parameters by
minimizing the difference between the $S/S_{ref}$ ratio and
the inverse sensitivity curve defined as the sum $\mathcal{C} + \mathcal{T}$.
The best fitted inverse sensitivity curve of an observation of the
calspec standard star BD+33d2642 is shown in fig.~\ref{fig:std_oct_stdstars}.

We store the four telluric parameters and the $\mathcal{C}$ inverse sensitivity spectrum,
assumed to be airmass-independent,
as a FITS file. When flux-calibrating a science target spectrum,
we use the target airmass rather than the standard star airmass in eq.~\ref{eq:telluric}.

\begin{figure}
  \centering
  \includegraphics[width=\linewidth]{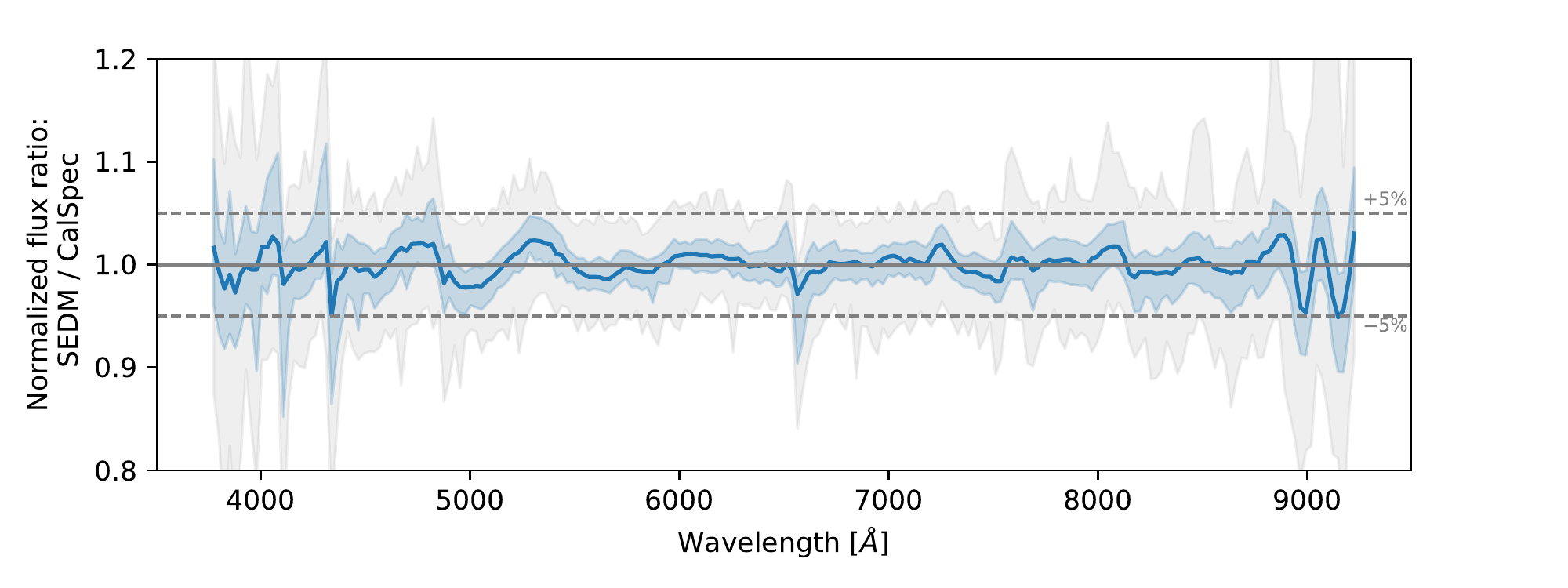}
  \caption{Mean-normalized ratio between flux-calibrated standard
    star spectra (considered as science targets, various stars have
    been used) and their expected CalSpec spectra. The solid blue 
    line shows the running median spectrum value, while the blue and
    grey bands encapsulate 68\% and 95\% of the spectra, respectively. 
    This figure illustrates that our color calibration is accurate at
    the few percent level.}
  \label{fig:fluxcal_fit}
\end{figure}

%
%
To test the quality of the flux calibration, we examine the ratio between
the reference spectrum and the flux-calibrated observed spectrum for any
standard star. In fig.~\ref{fig:fluxcal_fit},
we present the ratio between SEDM flux-calibrated standard star
spectra and their corresponding calspec spectra convolved to match the
SEDM wavelength scale. The flux ratios encapsulating 0.95 to 1.05 (pale
blue) and 0.84 to 1.16 (light gray) are indicated as well as the flux
ratio median (dark blue) for the 141 standard star observations taken in
October 2018 used for this test.  The plot shows the RMS scatter of the
median ratio is around 1\%, while the range of the 1$\sigma$ scatter stays
within an envelope of a few percent. 
It is therefore expected that our flux-calibrated science spectra are
\mr{relatively flux-calibrated} at the few percent level. A more
thorough analysis would be necessary 
for the spectrophotometric absolute calibration, but this is both beyond the scope of this paper
and the SEDM purpose in general. More generic tests of the
pipeline results will be presented in section~\ref{sec:results}.

\subsection{\mr{Accuracy of the wavelength calibration}}
\label{sec:test_wavesolution}

\begin{figure}
  \centering
  \includegraphics[width=\linewidth]{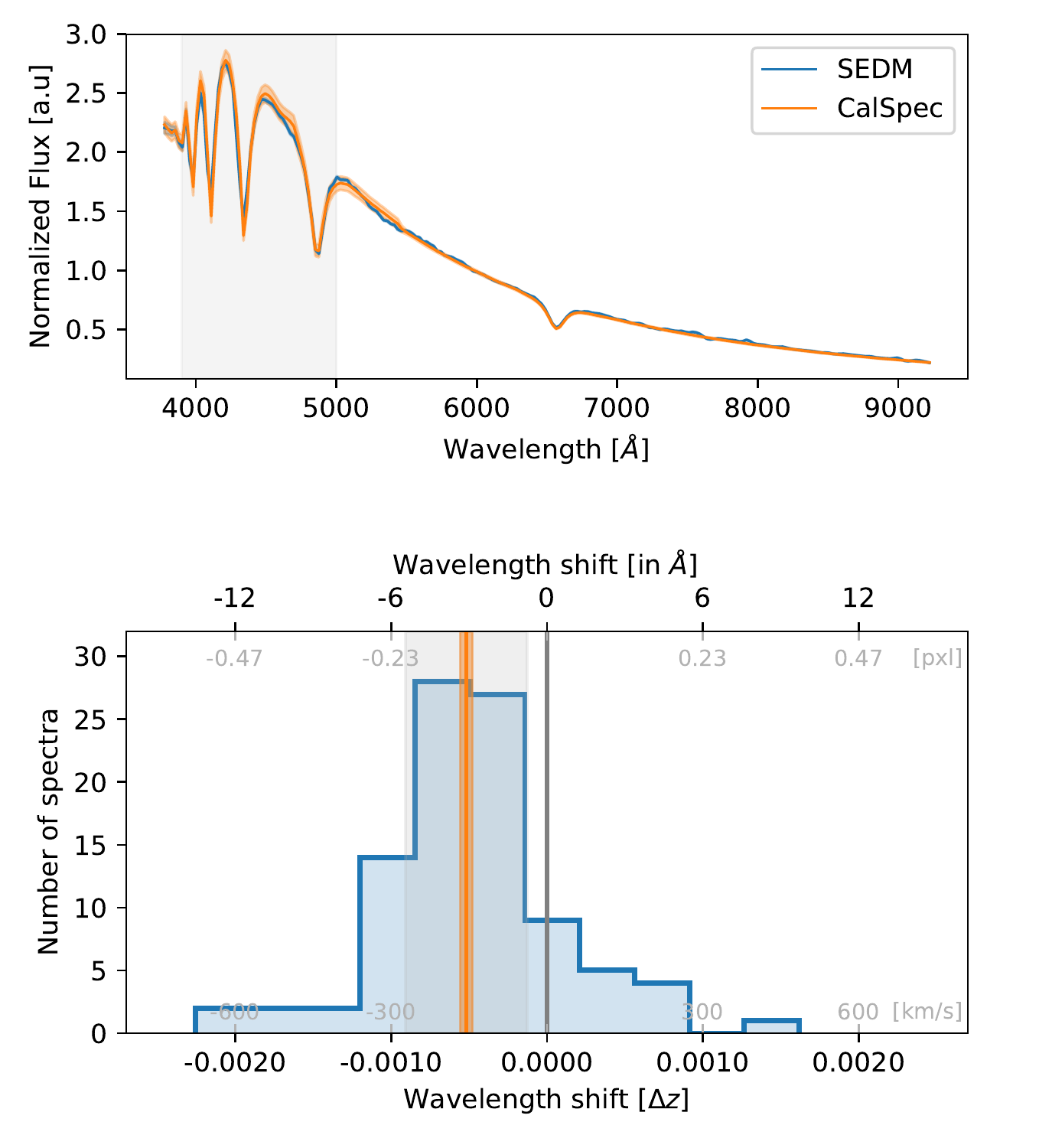}
  \caption{ \mr{Accuracy of the wavelength calibration. 
      \emph{Top}: A SEDM and a CalSpec spectrum of the HZ4 standard
      stars. The grey band shows the wavelength region used to fit for
      the wavelength shift between the two spectra. 
    \emph{Bottom}: Histogram of the wavelength shift fitted on the 94
    HZ4 star spectra obtained by SEDM in Septembre 2018. The lower
    axis presents the shift in redshift shift (in $z$ or in
    $\mathrm{km\,s^{-1}}$), the upper axis shows the corresponding
    offset in $\AA$ or pixel for a typical wavelength of 6000
    $\AA$. The vertical orange line represents the mean and error on
  the mean. The light-grey band shows the standard deviation. SEDM
  spectra automatically extracted by \pysedm{} have a precision of
  $\sim3\,\AA$ or $4\times10^{-4}$ in redshift. A small 1 std offset
  is visible suggesting that the SEDM spectra a lightly blue-shifted
  of approximately a tens of a pixel. The origin of this shift is
  under investigation.}}
  \label{fig:wavelength_cal}
\end{figure}

\mr{In section~\ref{sec:nightcalibration_wavesol} we presented the
construction of the nightly wavelength solution that is then fine
tuned for each science exposure to account for flexure ; see
section~\ref{sec:ccdtocube_iflexure}. In this section, we compare
\pysedm{} extraction of HZ standard star spectra with 
their corresponding CalSpec spectra. The upper panel of
fig~\ref{fig:wavelength_cal} shows an SEDM spectrum overplotted with a
CalSpec spectrum smoothed to match the SEDM wavelength resolution. As shown
in this figure, HZ stars have strong hydrogen absorption lines, which
enable us to test the wavelength accuracy of the \pysedm{}
pipeline. To do so, we fit the absorption line region [$3900,
5000\,\AA$] for a wavelength offset between the 
CalSpec spectrum and the SEDM spectrum together with a 2D (linear)
multiplicative polynomial to account for potential flux calibration
inaccuracy. We repeat this procedure for the 94 HZ4 star observed over
the one-month period in September, 2018. The lower panel of fig.~\ref{fig:wavelength_cal} shows the
histogram of the fitted wavelength shifts. The RMS of this distribution
is $\sim$$3\,\AA$, or $\sim$$0.1$ SEDM pixels. For a typical wavelength of
$6000\,\AA$ this corresponds to a precision of $4\times10^{-4}$ in redshift,
or $120\,\mathrm{km\,s^{-1}}$. Despite this high precision of the
\pysedm{} pipeline, figure~\ref{fig:wavelength_cal} shows a systematic
offset of $\sim3\,\AA$ toward the blue.  The origin and correction for this small, but significant, offset is under investigation.}

\subsection{Manual Extraction Option}
\label{sec:humain_interaction}
The pipeline is fully automated and needs no manual input from the data
acquisition to the generation of a flux-calibrated spectrum.
However, one can manually call any of the \pysedm\ functions to bypass some
of the automated steps.  In particular, the target position
and the part of the SEDM cube used to run the 3D PSF extraction can be
specified on the command line by simply providing
$\{x, y\}$ (the target reference centroid position guess otherwise given
by the astrometric procedure detailed in section~\ref{sec:guider}).
An interactive plot frontend also exists to manually
click on the expected target location and draw the extraction region.
Further details on how to interact with \pysedm\ is given in the online documentation\footnote{https://github.com/MickaelRigault/pysedm}.

\subsection{Quality flag, typing and check plots}
\label{sec:quality_control}
\begin{figure}
  \centering
  \includegraphics[width=\linewidth]{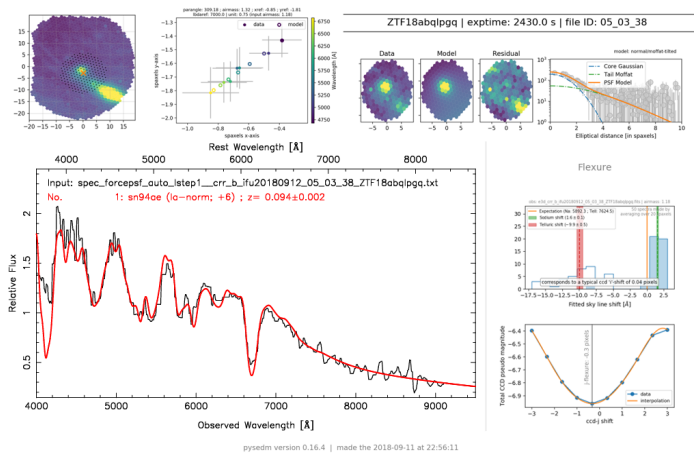}
  \caption{Summary figure presenting in one compact image the relevant
    plot automatically generated by the \pysedm{} pipeline. The figure
    is automatically created and pushed together with the final
    flux-calibrated spectrum to registered clients, notably the ZTF
    marshal. This example is a science observation of ZTF18abqlpgq.}
  \label{fig:pysedmreport_ZTF18abqlpgq}
\end{figure}

At the end of the automated processing, we apply the following criteria
to generate a quality flag value that is recorded in the output header
with the keyword QUALITY:
\begin{itemize}
\item 5: the automated target positioning in the MLA (see section~\ref{sec:guider}) failed,
\item 4: no object signal (defined as more than 20\% of the flux-calibrated spectrum is negative) 
\item 3: the automatic target positioning placed the target outside the MLA 
\item 0 for no problem
\end{itemize}

If the science target is not a standard star, and if its quality flag is 0,
we run SNID \citep{snid_2007} on the flux-calibrated spectrum 
to make an automated classification. If the SNID "rlap" of the first match is
greater than 4.0 we push this classification. We warn the user that
automatic classifications should be used with care.

At the end of each exposure, a summary plot encapsulating all the relevant
automatic extraction plots is saved and pushed to the SEDM clients,
including the ZTF Growth Marshal. We show in
fig.~\ref{fig:pysedmreport_ZTF18abqlpgq}
the "pysedm-report" automatically created for the observation of our example
Supernova ZTF18abqlpgq.
The SNID match is excellent with a Type Ia supernova at 6 days post-maximum at a redshift
of $z\sim0.094$.

\section{Results}
\label{sec:results}

\subsection{Some statistics for 19 weeks of operations}
\begin{figure}
  \centering
  \includegraphics[width=\linewidth]{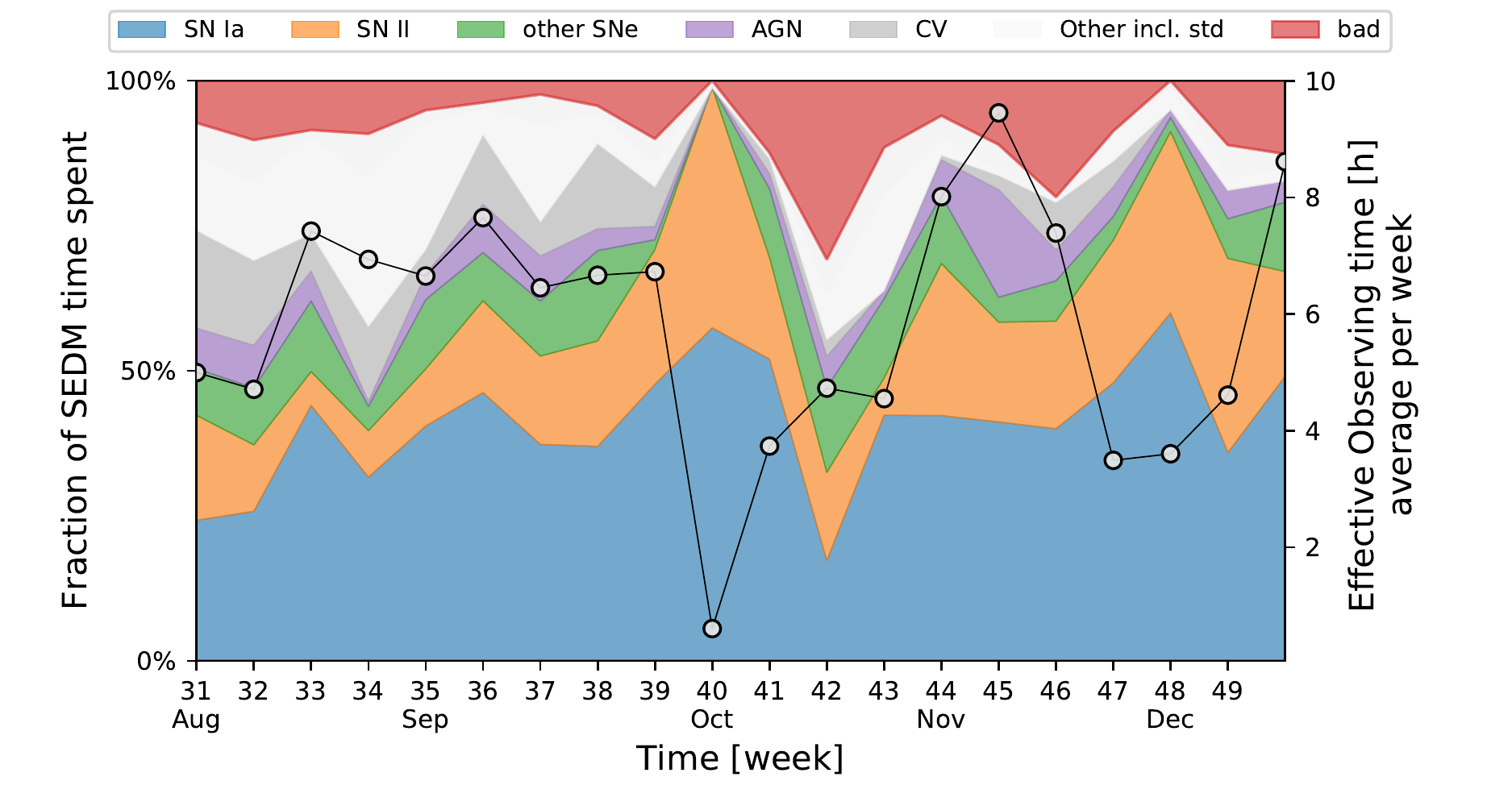}
  \caption{Use of the SEDM time since \pysedm{} is the sole SEDM
    pipeline (early august 2018) and until redaction of this paper (19
    weeks). 
    Color bands show the fraction of target observed per week sorted
    by type (see legend) ; left-axis. The running line with
    grey markers shows the evolution of the effective time spent on
    target per night (weekly average) ; right-axis.}
  \label{fig:sedm_timespent}
\end{figure}

Since the beginning of August 2018, \pysedm{} has been the sole pipeline for the SEDM. 
The ZTF collaboration is using this instrument to type the
transients discovered by the ZTF camera mounted on the Palomar 48"
Schmidt  (P48) telescope. We present in fig.~\ref{fig:sedm_timespent} the
fraction of object types generated by \pysedm{} as a function of time, and
the fraction of time spent acquiring SEDM observations. We spend, on average,
6h30m on targets each night. In addition, we have a typical overhead of 180s,
including slew time and readout. 
We point out two dips in the observing time; the deepest in mid-October and a
smaller dip in early December. These correspond to an engineering period
(mid-October), and a period of poor weather conditions, notably due to
wildfires in southern California, at the end of November to early December 2018.

About two thirds of the SEDM observing time is spent acquiring supernova spectra,
two thirds of which are Type Ia supernovae.
We see in fig.~\ref{fig:sedm_timespent} that the fraction of
time spent on standard stars (~90\% of ``other incl. std'') has
continuously been reduced from \textasciitilde 15\% to about 2\%. The reasons are
\mr{threefold}: 
first, we observed numerous standard stars when launching
\pysedm{} in production to asses the robustness of the pipeline ; 
second, at that time not all ZTF fields had references, which limited
the amount of SEDM triggering for transients ; \mr{and third, the
  human transient scanning has progressed optimizing the triggering of
  extra-galactic candidates}. 
 We also note that the fraction of
variable stars (``CV'') 
is reducing as a function of time. Except for
rare occasions, such targets are not observed on purpose, hence the
time spent observing them is reducing as ZTF triggering of
the SEDM machine is improving. Finally, one can notice that the
fraction of time spent acquiring spectra that turn out to be classified
as ``bad'' (quality $\neq$ 0, see section~\ref{sec:quality_control}) is
slightly increasing with time. This is due to observing conditions
getting worse while entering the winter period.

\begin{figure}
  \centering
  \includegraphics[width=\linewidth]{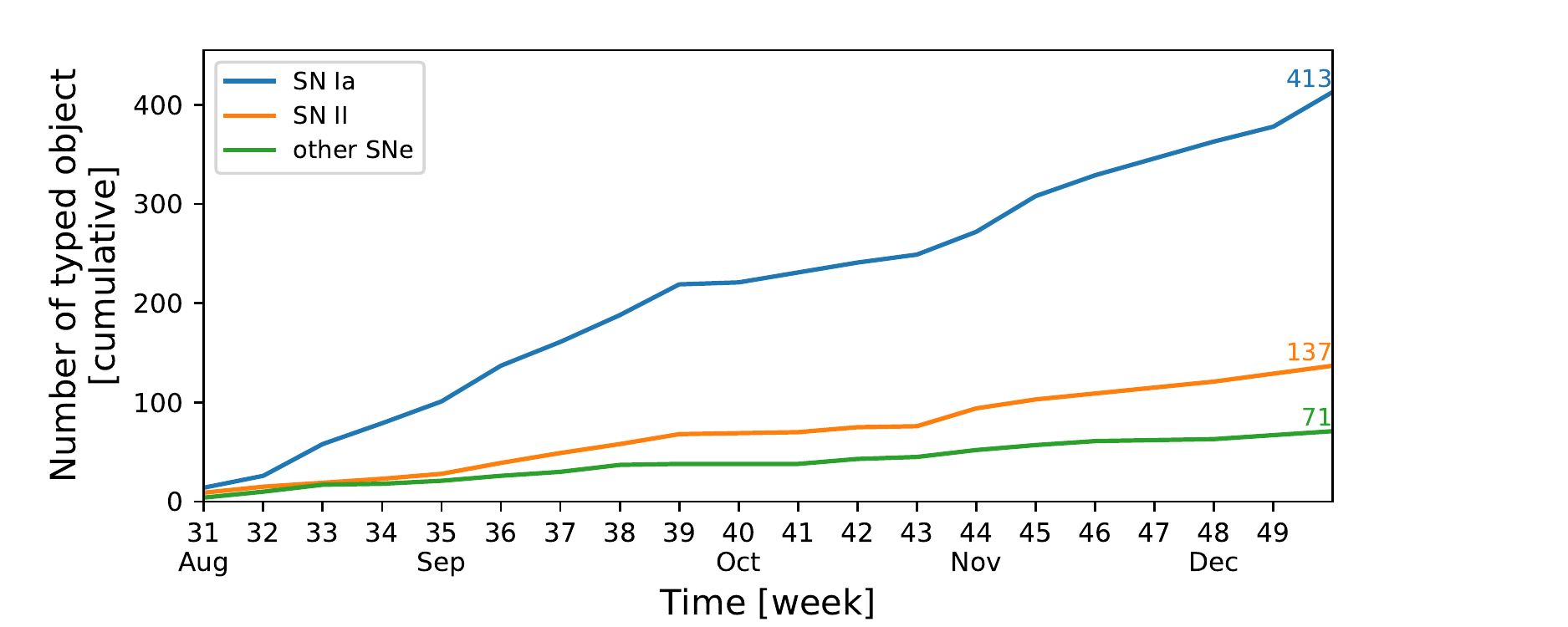}
  \caption{Number of supernovae typed by the SEDM as a function of time.}
  \label{fig:sedm_typing}
\end{figure}

Figure~\ref{fig:sedm_typing} presents the number of individual supernovae
typed by the SEDM since the launch of \pysedm{} in early August. 
In 19 weeks, we have typed \textasciitilde  400 SNe Ia, almost \textasciitilde  140 SNe II
and \textasciitilde  70 other SNe, mainly Ib/c. This corresponds to
roughly 20 new SNe Ia and 7 new Type II per week. \mr{See classification
details and more accurate discovery estimation in Fremling et al 
  (in prep)}


\subsection{Spectral accuracy: comparison with other spectrographs}
\label{sec:comparison}

\begin{figure*}
  \centering
  \includegraphics[width=\linewidth]{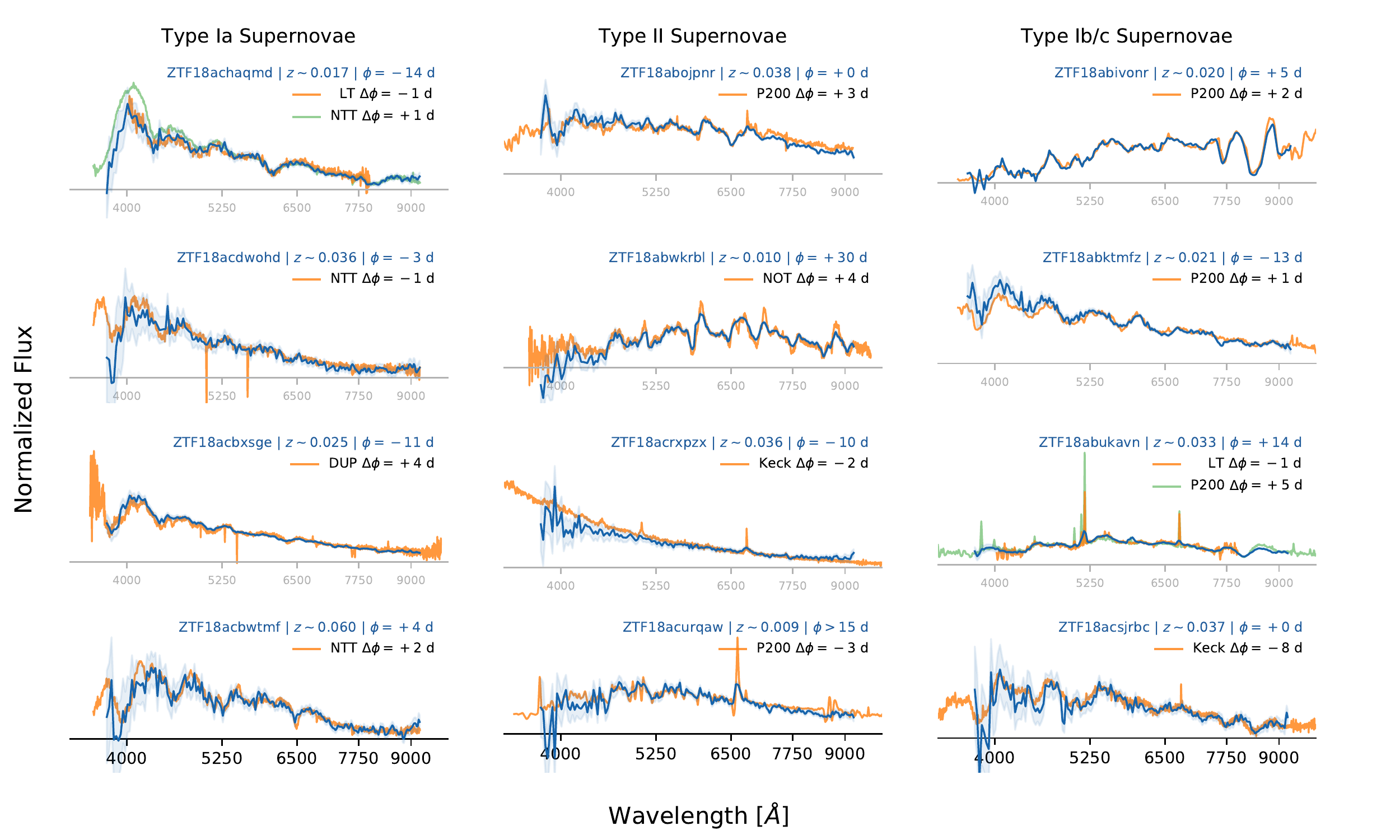}
  \caption{Comparison between SEDM spectra (in blue) with other higher
    resolution instruments. Legends in each panel present: in blue
    the target name, redshift and phase of the SEDM spectrum; in color (orange and/or
    green), the other facility name and the phase difference with
    respect to the SEDM spectrum. \emph{Facility names}: 
    NTT \citep[New Technology Telescope, ePESSTO,
    see][http://www.pessto.org]{smartt_pessto_2015}, 
    LT \citep[Liverpool Telescope JMU; SPRAT][]{sprat_paper}, 
    DUP (LCO-duPont ; WFCCD), 
    P200 (Palomar 200-inch Telescope; Double Spectrograph), 
    Keck (Keck Telescope; LRIS), NOT (Nordic Optical Telescope ;
    Alhambra Faint Object Spectrograph and Camera). For visibility
    P200 spectra 
    have been smoothed by a five-bin Gaussian filter.}
  \label{fig:sedm_and_others}
\end{figure*}

In some (rare) instances, a target observed by the SEDM has also been
observed by another spectroscopic facility at close to the same
time, typically with higher spectroscopic resolution. We use this
opportunity to test the quality of the \pysedm{}
spectral extraction. In Fig~\ref{fig:sedm_and_others}, we show four representative spectral
comparisons for each of three supernova types: Type Ia, Type II and
Type Ib/c. 
The objects were selected to be representative of our
sample spanning different phases, redshifts and magnitudes. The
characteristics of these spectra are summarized in
Table~\ref{tab:supernova_sedm}.  

This figure demonstrates the quality of the SEDM spectra
extracted using \pysedm{}. For each case, the overall shape matches
that of the other spectrum or spectra, which
confirms the accuracy of the color calibration as discussed in 
section~\ref{sec:fluxcal}. In addition, the main characteristic
features, such as the silicon line at~6500~$\AA$ for the SNe~Ia, or the
P-Cygni H$\alpha$ shape for the Type II, are clearly visible in the
SEDM spectra. This shows that a very low-resolution
spectrograph is well suited for transient typing \citep{sedm_paper}.  
However, galaxy emission lines, as seen in the spectra of ZTF18abukavn,
are not well resolved by the SEDM \mr{(see details concerning this
target in Ho et al. in prep)}. This could limit our ability to
accurately derive host redshifts from SEDM IFU data as done, e.g. with
SNIFS \citep{rigault_2013,rigault_2018}; investigation into improvement
is ongoing.

\begin{table*}
\centering
\caption{Summary of supernovae presented in fig.~\ref{fig:sedm_and_others}.
}
\label{tab:supernova_sedm}
\begin{tabular}{l | c c c c c c c c}
\hline\hline\\[-0.8em]
ZTF name  & TNS name & RA & Dec & redshift & typing & phase & magnitude & Exptime \\[0.15em]
                 &                  & deg & deg  &            &  & days  & ``ztf:g''& s \\[0.15em]
\hline\\[-0.8em]
ZTF18acbwtmf & SN2018hrn & 322.8909 & +23.0147 & 0.060 & Ia & 4 & 18.1 & 1200\\[0.30em]
ZTF18acbxsge & SN2018huw & 9.8009 & -5.2007 & 0.025 & Ia & -11 & 17.6 & 1600\\[0.30em]
ZTF18acdwohd & SN2018ids & 319.0474 & +11.9462 & 0.036 & Ia & -3 & 16.5 & 1600\\[0.30em]
ZTF18achaqmd & SN2018ilu & 353.3374 & +4.8096 & 0.017 & Ia & -14 & 17.2 & 1200\\[0.30em]
ZTF18acurqaw & SN2018hwm & 125.3675 & +3.1645 & 0.009 & II & >15 & 19.0 & 2250\\[0.30em]
ZTF18acrxpzx & SN2018jrj & 167.7667 & +64.2474 & 0.036 & II & -10 & 18.3 & 2250\\[0.30em]
ZTF18abwkrbl & SN2018gjx & 34.0649 & +28.5913 & 0.010 & IIb & 30 & 18.5 & 1440\\[0.30em]
ZTF18abojpnr & SN2018fzn & 297.4871 & +59.5928 & 0.037 & IIb & 2 & 19.3 & 2430\\[0.30em]
ZTF18acsjrbc & SN2018jme & 96.1574 & +50.9750 & 0.037 & Ic &  0 & 19.1 & 2250\\[0.30em]
ZTF18abukavn & SN2018gep & 250.9509 & +41.0454 & 0.033 & Ic-BL & 12 & 17.9 & 1440\\[0.30em]
ZTF18abktmfz & SN2018eoe & 336.1015 & +18.0719 & 0.020 & Ib & -13 & 18.8 & 2430\\[0.30em]
ZTF18abivonr &      --          & 321.8796 & +45.4223 & 0.020 & Ic & 5 & 19.4 & 2430\\[0.30em]
\hline
\end{tabular}
\tablefoot{
phases and magnitudes are approximations derived from light curve fitting
using sncosmo. They are illustrative and should be used with care,
especially for Type II supernovae.}  
\end{table*}

\section{Summary and conclusions}
\label{sec:conclusion}

This paper presents \pysedm{}: a fully automated pipeline for the
SEDMachine \citep{sedm_paper}.  First, we detailed the procedure for
constructing an $x,y,\lambda$ 3D-cube from the 2D CCD exposures acquired by
the integral field unit. We presented the construction of the
wavelength solution, the trace identification, the flat fielding, and
the flexure corrections. 
Then, we detailed the automatic target point source spectral extraction
from the 3D cube. For this step, based on 3D PSF modeling, we
first-guess the target location in the IFU by projecting the WCS
solution measured on guider images taken simultanesously. 
Following \cite{buton_2013}, The PSF model is a linear combination of
Moffat and Gaussian elliptical profiles.
Altogether, the \pysedm{} pipeline enables a fully automated spectral 
extraction of a targeted point source object within 5 minutes after the
end of exposure. This spectrum is typed using SNID
and pushed to registered clients, notably the ZTF marshal. If the SNID
typing quality value rlap is higher than 4, the SNID type is also pushed.

While no human interaction is required, \pysedm{} conveniently allows
manual re-extraction or fine-tuning of any step of the pipeline. For instance,
one can fine-tune the target location and/or the IFU area
considered for 3D PSF extraction in order to avoid bright source
contamination. 

We show in this paper that SEDM spectra automatically extracted by
\pysedm{} are \mr{relatively flux-calibrated} at the few percent level,
which enables spectroscopic typing. Since \pysedm{} has entered
production in early
August, we have, \mr{among other objects} classified about 400 Type Ia
supernovae and  140 Type II supernovae in 19 weeks \mr{; see
  classification details in Fremling et al (in prep).} 

To conclude, we would like to emphasize the power of having fully 
automated spectroscopic facilities in the era of modern time-domain imaging
surveys such as ZTF today and LSST in the near future. LSST will
operate in the south and will discover up to ten times more
transients than ZTF. A fraction of those will be near
enough in redshift (say z<0.2) that one could imagine getting a
spectrum of most, if not all of them with just a few SEDM-like facitities
installed on 2-m class telescopes in the southern hemisphere. \pysedm{}
has been developed to be extremely flexible and could therefore be
adapted to any IFU-like spectrograph with very limited effort. Also,
while \mr{the SEDM very low-resultion $(R\sim100)$ is great for
typing,} from the perspective of developing new spectroscopic
follow-up 
facilities, we would encourage the use of a  higher wavelength
resolution design \mr{($R\sim1000$), high enough} to resolve the
H$\alpha$ [NII] structure in order to accurately enable host
spectroscopic redshift measurements simultaneously with the transient
typing.

\begin{acknowledgements}

\mr{We thank D. O. Cook, K. De, K. B. Burdge and A. Y. Ho for generously
sharing their spectra for the ``SEDM and other spectrograph'' figure.}
This project has received funding from the European Research Council
(ERC) under the European Union's Horizon 2020 research and innovation
program (grant agreement n$^\circ$759194 - USNAC).
The SED Machine is based upon work supported by the National Science
Foundation under Grant No. 1106171. 
Based on observations obtained with the Samuel Oschin Telescope
48-inch and the 60-inch Telescope at the Palomar Observatory as part
of the Zwicky Transient Facility project. ZTF is supported by the
National Science Foundation under Grant No. AST-1440341 and a
collaboration including Caltech, IPAC, the Weizmann Institute for
Science, the Oskar Klein Center at Stockholm University, the
University of Maryland, the University of Washington, Deutsches
Elektronen-Synchrotron and Humboldt University, Los Alamos National
Laboratories, the TANGO Consortium of Taiwan, the University of
Wisconsin at Milwaukee, and Lawrence Berkeley National
Laboratories. Operations are conducted by COO, IPAC, and UW.

\end{acknowledgements}


\end{document}